\documentclass{aa}
\usepackage[varg]{txfonts}

\usepackage{placeins}
\usepackage[colorlinks=true, linkcolor=blue, urlcolor=blue, citecolor=blue, anchorcolor=blue]{hyperref}
            
\usepackage{natbib}
\bibpunct{(}{)}{;}{a}{}{,} 

\title{The ESO SupJup Survey I: Chemical and isotopic characterisation of the late L-dwarf DENIS J0255-4700 with CRIRES$^+$}

\titlerunning{Chemical and isotopic characterisation of a late L-dwarf with CRIRES$^+$}

\author{
S. de Regt\inst{\ref{inst1}} \and
S. Gandhi\inst{\ref{inst1},\ref{inst2},\ref{inst3}} \and
I. A. G. Snellen\inst{\ref{inst1}} \and
Y. Zhang\inst{\ref{inst1},\ref{inst4}} \and
C. Ginski\inst{\ref{inst5}} \and
D. Gonz\'alez Picos\inst{\ref{inst1}} \and
A. Y. Kesseli\inst{\ref{inst6}} \and
R. Landman\inst{\ref{inst1}} \and
P. Molli\`ere\inst{\ref{inst7}} \and
E. Nasedkin\inst{\ref{inst7}} \and
A. S\'anchez-L\'opez\inst{\ref{inst8}}\and
T. Stolker\inst{\ref{inst1}}
}

\institute{
Leiden Observatory, Leiden University, P.O. Box 9513, 2300 RA, Leiden, The Netherlands\\\email{regt@strw.leidenuniv.nl}\label{inst1} \and
Department of Physics, University of Warwick, Coventry CV4 7AL, UK\label{inst2} \and
Centre for Exoplanets and Habitability, University of Warwick, Gibbet Hill Road, Coventry CV4 7AL, UK\label{inst3} \and
Department of Astronomy, California Institute of Technology, Pasadena, CA 91125, USA\label{inst4} \and
School of Natural Sciences, Center for Astronomy, University of Galway, Galway, H91 CF50, Ireland\label{inst5} \and
IPAC, Mail Code 100-22, Caltech, 1200 E. California Boulevard, Pasadena, CA 91125, USA \label{inst6} \and
Max-Planck-Institut für Astronomie, Königstuhl 17, 69117 Heidelberg, Germany \label{inst7} \and
Instituto de Astrof{\'i}sica de Andaluc{\'i}a (IAA-CSIC), Glorieta de la Astronom{\'i}a s/n, 18008 Granada, Spain\label{inst8}
}

\date{Received date /
Accepted date }

\abstract
{It has been proposed that the distinct formation and evolutionary pathways of exoplanets and brown dwarfs may affect the chemical and isotopic content of their atmospheres. Recent work has indeed shown differences in the $\mathrm{^{12}C/^{13}C}$ isotope ratio, provisionally attributed to the top-down formation of brown dwarfs and the core accretion pathway of super-Jupiters.} 
{The ESO SupJup Survey is aimed at disentangling the formation pathways of isolated brown dwarfs and planetary-mass companions using chemical and isotopic tracers. The survey utilises high-resolution spectroscopy with the recently upgraded CRyogenic high-resolution InfraRed Echelle Spectrograph (CRIRES$^+$) at the Very Large Telescope, covering a total of 49 targets. Here, we present the first results of this survey: an atmospheric characterisation of DENIS J0255-4700, an isolated brown dwarf near the L-T transition.}
{We analyse its observed CRIRES$^+$ K-band spectrum using an atmospheric retrieval framework where the radiative transfer code \texttt{petitRADTRANS} is coupled with the \texttt{PyMultiNest} sampling algorithm. Gaussian Processes are employed to model inter-pixel correlations. In addition, we adopt an updated parameterisation of the pressure-temperature profile.}
{Abundances of CO, H$_2$O, CH$_4$, and NH$_3$ are retrieved for this fast-rotating L-dwarf. The ExoMol H$_2$O line list provides a significantly better fit than that of HITEMP. A free-chemistry retrieval is strongly favoured over equilibrium chemistry, caused by an under-abundance of CH$_4$. The free-chemistry retrieval constrains a super-solar $\mathrm{C/O}$-ratio of $\sim$\,$0.68$ and a solar metallicity. We find tentative evidence ($\sim$\,$3\sigma$) for the presence of $^{13}$CO, with a constraint on the isotopologue ratio of $\mathrm{^{12}CO/^{13}CO}=184^{+61}_{-40}$ and a lower limit of $\gtrsim$\,$97$, which suggests a depletion of $\mathrm{^{13}C}$ compared to the local interstellar medium ($\mathrm{^{12}C/^{13}C}\sim$\,$68$).}
{High-resolution, high signal-to-noise K-band spectra provide an excellent means to constrain the chemistry and isotopic content of sub-stellar objects, as is the main objective of the ESO SupJup Survey.}


\usepackage{float}
\usepackage{subfiles}
\usepackage{pdflscape}
\usepackage{xcolor}
\usepackage{xurl}

\begin{document}

\maketitle

\section{Introduction}
Spectroscopic observations can be used to constrain the chemical composition, thermal and cloud structure, and dynamics of exoplanet atmospheres. It has been proposed that spectral characterisation of the chemistry of exoplanet atmospheres can help to shed light on planet formation and evolutionary processes. The chemical make-up of the solid and gaseous planetary building blocks is expected to be set by various processes that depend on the location in the disk (e.g. \citealt{Alarcon_ea_2020,Turrini_ea_2021,Pacetti_ea_2022,Molliere_ea_2022}).
Therefore, a number of chemical abundance ratios have been suggested as tracers of planet formation and evolution, in particular the carbon-to-oxygen ratio ($\mathrm{C/O}$; \citealt{Oberg_ea_2011,Madhusudhan_ea_2012,Mordasini_ea_2016}), nitrogen-to-oxygen or nitrogen-to-carbon ratio ($\mathrm{N/O}$, $\mathrm{N/C}$; \citealt{Cridland_ea_2016,Turrini_ea_2021}), and the refractory-to-volatile ratio \citep{Lothringer_ea_2021}. 

Additionally, isotope ratios have been proposed as complementary tracers of planet histories \citep{Clayton_ea_2004,Molliere_ea_2019a,Zhang_ea_2021a,Zhang_ea_2021b}. In the Solar System, for instance, the measured deuterium-to-hydrogen (D/H) ratios of Uranus and Neptune display an enhancement of deuterium by about a factor of 2 compared to the proto-solar abundance \citep{Feuchtgruber_ea_2013}. This discrepancy is suggested to be caused by the accretion of HDO-rich ices. Moreover, the increased D/H ratios of Mars and Venus are indicative of atmospheric losses \citep{Kulikov_ea_2006,Villanueva_ea_2015,Alday_ea_2021}, where the lighter isotopologue is more readily removed. While the $\mathrm{^{12}C/^{13}C}$ ratio shows minimal variation in the Solar System \citep{Woods_ea_2009}, recent measurements for exoplanet atmospheres \citep{Zhang_ea_2021a,Line_ea_2021,Gandhi_ea_2023b} have highlighted the potential of the carbon isotope ratio to serve as an additional diagnostic of planet formation histories. \citet{Zhang_ea_2021a} detected the $\mathrm{^{13}CO}$ isotopologue in the atmosphere of the young, super-Jupiter YSES 1b. Using an atmospheric retrieval analysis, the $\mathrm{^{12}CO/^{13}CO}$ abundance ratio was determined to be $\sim$\,$31$. Hence, the atmosphere of YSES 1b appears to be significantly enriched with $\mathrm{^{13}C}$ compared to the carbon isotope ratio of the local interstellar medium (ISM; $\mathrm{^{12}C/^{13}C}\sim$\,$68$; \citealt{Langer_ea_1993,Milam_ea_2005}). The accretion of $\mathrm{^{13}C}$-rich ices beyond the CO snowline was put forward as an explanation of this discrepancy. Observations of a young, isolated brown dwarf (2M 0355) revealed an isotopologue ratio of $\mathrm{^{12}CO/^{13}CO}\sim$\,$108$ \citep{Zhang_ea_2021b,Zhang_ea_2022}. The different isotope ratios of the brown dwarf and super-Jupiter could be a sign of their distinct formation pathways. The brown dwarf is thought to form via the gravitational collapse of a gas cloud, whereas the super-Jupiter possibly forms via core-accretion which can affect its isotopic composition depending on the fractionation processes in the protoplanetary disk \citep{Zhang_ea_2021b}. 

In this paper, we present an atmospheric retrieval analysis of the K-band spectrum of the isolated brown dwarf DENIS J025503.3-470049 (hereafter: DENIS J0255). This spectrum was observed with the upgraded CRyogenic high-resolution InfraRed Echelle Spectrograph (CRIRES$^+$) as part of the ESO SupJup Survey. Section \ref{sect:SupJup} introduces the ESO SupJup Survey and describes the observed low- and planetary-mass objects. In Sect. \ref{sect:DENIS_J0255}, we introduce DENIS J0255 as the target of this study. Furthermore, we explain the reduction of its spectral observations and the retrieval framework employed in the analysis. Section \ref{sect:results} details the results of the retrieval analysis. Finally, Sect. \ref{sect:conclusions} summarises the conclusions of this study.

\section{ESO SupJup Survey}\label{sect:SupJup}
The ESO SupJup Survey (Program ID: 1110.C-4264, PI: Snellen) is aimed at disentangling the formation pathways of a sample of super-Jupiters, free-floating planets and brown dwarfs. For these low- and planetary-mass objects, we wish to constrain the (thermal) atmospheric structures, chemical abundances, surface gravities, possible accretion signatures, and rotation velocities. A particular objective is to constrain the $^{12}$C/$^{13}$C isotope ratio, in combination with the C/O-ratio and metallicity, all proposed as tracers of the formation histories of sub-stellar objects \citep{Molliere_ea_2019a,Zhang_ea_2021b}. 
Recent work has established high-resolution spectroscopy \citep{Birkby_2018} as an effective technique to infer the presence of molecular and atomic species (e.g. \citealt{Brogi_ea_2012,Hoeijmakers_ea_2020,Giacobbe_ea_2021}), their (relative) abundances (e.g. \citealt{Brogi_ea_2017,Line_ea_2021,Gandhi_ea_2023a}), as well as the planet rotation (e.g. \citealt{Snellen_ea_2014,Schwarz_ea_2016}) and atmospheric wind velocities (e.g. \citealt{Snellen_ea_2010,Brogi_ea_2016,Ehrenreich_ea_2020}).

\begin{figure}[h!]
    \resizebox{\hsize}{!}{\includegraphics[width=17cm]{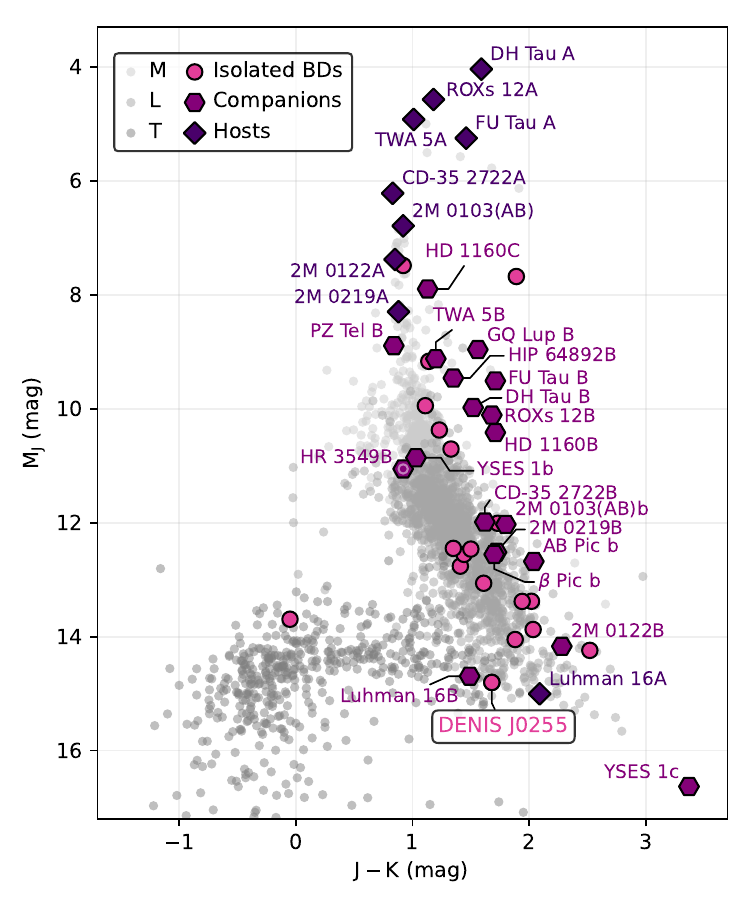}}
    \caption{Colour-magnitude diagram displaying the diverse sample of low- and planetary-mass objects observed as part of the ESO SupJup Survey. The pink, circular markers indicate the observed isolated brown dwarfs. The purple hexagons and dark purple diamonds depict the observed companions and their hosts, respectively. As a reference, the photometry of isolated brown dwarfs was obtained from the UltracoolSheet (\url{http://bit.ly/UltracoolSheet}) and is used to display late M, L and T dwarfs with increasingly darker marker shades. HR 3549B shows the $\mathrm{J}-\mathrm{H}$ colour as its K-band magnitude has not been measured.}
    \label{fig:CMD}
\end{figure}

\begin{figure}[h!]
    \resizebox{\hsize}{!}{\includegraphics[width=17cm]{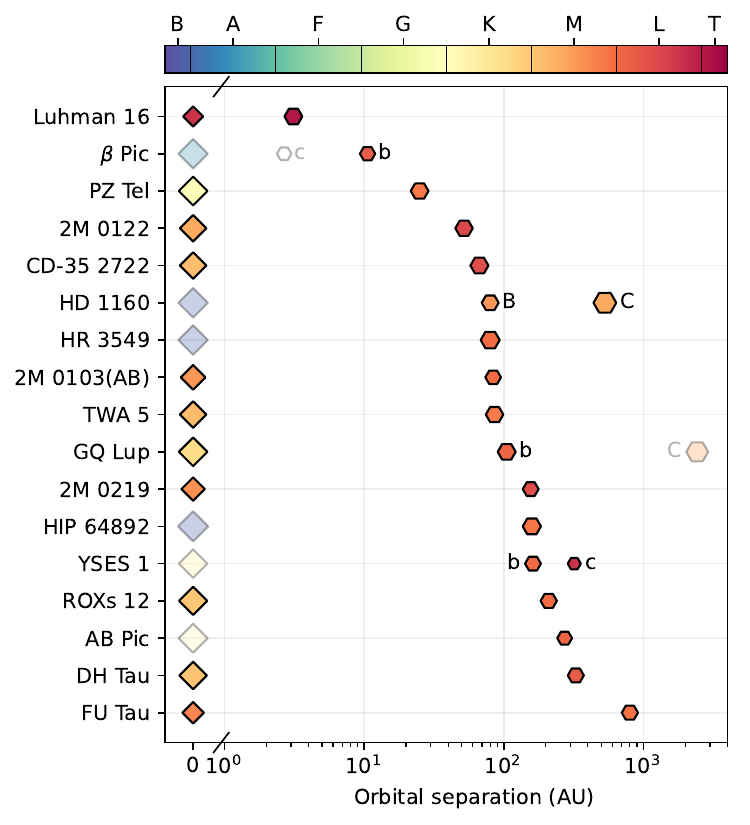}}
    \caption{Orbital configuration of the 17 observed bound systems. The systems are ordered by orbital separation (x-axis) of the closest-in observed companion. Marker colours and sizes indicate the spectral types and approximate masses. $\beta$ Pic c is not filled-in since its spectral type is unknown. Transparent symbols denote objects that were not centred on the slit.}
    \label{fig:separation}
\end{figure}

The SupJup Survey was carried out with the recently upgraded CRIRES$^+$ spectrograph \citep{Kaeufl_ea_2004,Dorn_ea_2014,Dorn_ea_2023}, installed at the Very Large Telescope (UT3: Melipal) in Chile. Over 14 nights, distributed in runs from November 2022 to March 2023, we obtained high-resolution spectra of 19 isolated objects, 19 lower-mass companions, and 11 hosts. Table \ref{tab:observed_targets} summarises the basic properties of the observed objects reported in the literature, and the utilised observing strategies. The sample of sub-stellar objects was selected in such a way that it covers diverse spectral types, as is illustrated in the colour-magnitude diagram of Fig. \ref{fig:CMD}. The isolated objects, shown with pink circular markers, cover the range from mid-M dwarfs (M6; 2MASS J12003792-7845082) to mid-T dwarfs (T4.5; 2MASS J05591914-1404488). The observed companions (hexagonal markers) range from mid-M dwarfs (M3.5; HD 1160C) to early-T dwarfs (T0.5; Luhman 16B), whereas the hosts (diamond markers) mostly consist of M dwarfs, with the exception of Luhman 16A (L7.5). The planetary-mass companions exhibit redder colours compared to their isolated, field counterparts (grey markers in Fig. \ref{fig:CMD}) which is likely the result of a cloudier atmosphere caused by their lower surface gravities \citep{Saumon_ea_2008,Marley_ea_2012}. In general, observations were taken in the K2166 wavelength setting, in order to cover the $^{13}$CO bandhead near $\sim$\,$2.345\ \mathrm{\mu m}$. For some objects spectra were also obtained in the J1226 wavelength setting (see Table \ref{tab:observed_targets}). These J-band spectra cover the K I and Na I alkali lines in addition to absorption from FeH, which are sensitive to the surface gravity \citep{McGovern_ea_2004,Allers_ea_2013}. 

Besides diversity in spectral type, planetary-mass companions were observed with a wide range of orbital separations, as is illustrated in Fig. \ref{fig:separation}, showing the orbital configuration of the 17 observed multi-object systems. Spectral type and approximate mass are indicated by colour and marker size, respectively. Off-slit hosts or companions are left transparent. We note that some off-slit hosts (e.g. $\beta$ Pic A, HD 1160A) have halos bright enough that their spectra are still observed at the chosen slit positions. Two orders of magnitude are covered in orbital separation, ranging from $\sim$\,$3\ \mathrm{AU}$ (Luhman 16B; \citealt{Luhman_2013}) to $\sim$\,$800\ \mathrm{AU}$ (FU Tau B; \citealt{Luhman_ea_2009}). The wide range in orbital separations can potentially highlight isotopic diversity since the carbon isotope ratio is expected to depend on the location of a planet's formation within the circumstellar disk \citep{Zhang_ea_2021a}. Similarly, the observations of multiple companions in the HD 1160 and YSES 1 systems can help to constrain their joint formation scenarios. 

\section{DENIS J0255} \label{sect:DENIS_J0255}
As a consequence of the similar bulk properties to super-Jupiters, brown dwarfs provide an opportunity to study the atmospheres of planetary-mass objects with high signal-to-noise ratios due to their proximity to Earth and the absence of contamination from a stellar host. As such, constraints on the atmospheric and chemical properties of brown dwarfs can serve as benchmarks for the study of giant exoplanet atmospheres. Additionally, high-resolution spectra of brown dwarfs and low-mass stars can be used to validate the accuracy and completeness of molecular line lists (e.g. \citealt{Kesseli_ea_2020,Cont_ea_2021,de_Regt_ea_2022,Tannock_ea_2022}).

As brown dwarfs age and cool, their atmospheres are subjected to physical and chemical evolution that can be observed in the emergent spectra, colours and magnitudes \citep{Marley_ea_2010,Charnay_ea_2018}. A strong transition occurs between the hotter L-type dwarfs and the cooler T-dwarfs. The late L-dwarfs exhibit redder colours, as seen in the colour-magnitude diagram of Fig. \ref{fig:CMD}, and present CO absorption in their spectra. At lower temperatures, the T-dwarfs appear bluer and CH$_4$ becomes the dominant carbon-bearing molecule in the observed emission spectra \citep{Cushing_ea_2005}. The presence of clouds has commonly been invoked to explain the L-T transition \citep{Allard_ea_2001,Ackerman_ea_2001,Burrows_ea_2006,Saumon_ea_2008}. These cloud layers would form in the photosphere for L-dwarfs, but recede below it as the temperature decreases towards T-type dwarfs. Alternatively, recent work has suggested that chemical convection could give rise to the transition \citep{Tremblin_ea_2015,Tremblin_ea_2016}. Studying objects near the L-T transition can help to better understand its origin. As such, the field L9-dwarf DENIS J025503.3-470049 offers a prime opportunity to examine the atmospheric processes at the bottom of the L-dwarf branch \citep{Martin_ea_1999,Burgasser_ea_2006}. The effective temperature of DENIS J0255 is expected to be $T_\mathrm{eff}\sim$\,$1400\ \mathrm{K}$ and its spectrum exhibits signs of a high surface gravity $\log\textit{g}\gtrsim$\,$5$ \citep{Cushing_ea_2008,Tremblin_ea_2016,Charnay_ea_2018,Lueber_ea_2022}, anticipated for a relatively old brown dwarf ($2$--$4\ \mathrm{Gyr}$ from kinematic arguments; \citealt{Creech-Eakman_ea_2004}). Notably, spectroscopic studies at low- and moderate resolution have revealed the presence of CH$_4$ absorption \citep{Cushing_ea_2005,Roellig_ea_2004} as well as shallow absorption around $\sim$\,$11\ \mathrm{\mu m}$ which could be attributed to both an NH$_3$ \citep{Creech-Eakman_ea_2004} and a silicate cloud feature \citep{Roellig_ea_2004}. Measurements at higher resolving powers revealed a high projected rotational velocity of $\textit{v}\sin\textit{i}\sim$\,$40\ \mathrm{km\ s^{-1}}$ \citep{Basri_ea_2000,Mohanty_ea_2003,Zapatero_Osorio_ea_2006}. Unlike stars, mature brown dwarfs likely retain high rotation rates due to the reduced loss of angular momentum via magnetic winds \citep{Reiners_ea_2008,Bouvier_ea_2014}.

\subsection{Observations \& reduction} \label{sect:obs_and_reduction}
DENIS J0255 was observed on November 2nd, 2022 as part of the ESO SupJup survey (Program ID: 110.23RW.001, PI: Snellen). The observations were performed in 3 ABBA nod-cycles, resulting in 12 exposures of 300 seconds each. As a consequence of its faint R-band magnitude ($\mathrm{R}\sim$\,$19.9\ \mathrm{mag}$; \citealt{Costa_ea_2006}), Adaptive Optics (AO) could not be used. The K2166 wavelength setting was chosen in order to cover the $^{13}\mathrm{CO}$ bandhead near $2.345\ \mathrm{micron}$ ($\nu=2-0$ transition). The Differential Image Motion Monitor (DIMM) malfunctioned during the observations, but recorded an optical seeing of $\sim$\,$0.7"$ and $0.4"$ before and after this intermission. These seeing measurements are reasonable assumptions during the observations since the slit viewer camera showed a stable point-spread function. As the seeing was sufficiently low, the spectra were observed with the $0.2"$ slit to attain the highest spectral resolution ($R=\lambda/\Delta\lambda\sim$\,$100\,000$). Prior to observing DENIS J0255, observations were made of a telluric standard star, kap Eri, using the same slit, wavelength setting and also without AO. A single ABBA nod-cycle with 15-second exposures resulted in a signal-to-noise of $\sim$\,$240$ per pixel near $2.345\ \mathrm{\mu m}$, which was deemed sufficient to correct for the absorption from the Earth's atmosphere. The telluric absorption lines were also used to perform a secondary wavelength correction. 

\begin{figure*}[h!]
    \centering
    \includegraphics[width=17cm]{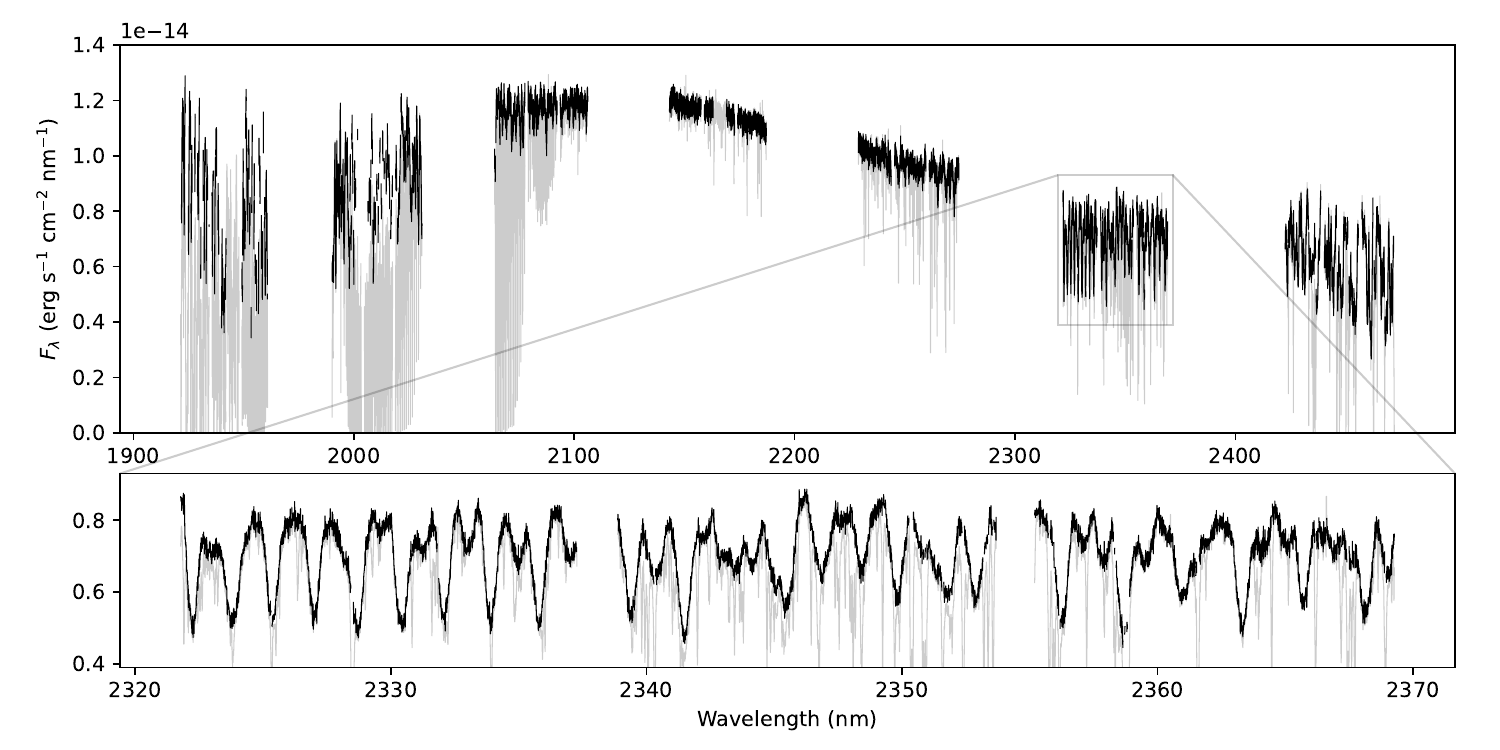}
    \caption{Calibrated spectrum of DENIS J0255, observed with CRIRES$^+$ in the K2166 wavelength setting. The y-axes indicate the flux $F_\mathrm{\lambda}$ in units of $\mathrm{erg\ s^{-1}\ cm^{-2}\ nm^{-1}}$ and the x-axes denote the wavelength $\lambda$ in $\mathrm{nm}$. \textit{Top panel}: spectrum over the full wavelength range, with the black line indicating the spectrum after the calibration described in Sect. \ref{sect:obs_and_reduction}. The grey line shows the observed spectrum without correcting for telluric absorption and removing outliers. \textit{Bottom panel}: zoom-in of the 6th spectral order, presenting numerous rotationally broadened spectral lines of $^{12}\mathrm{CO}$.}
    \label{fig:observed_spectrum}
\end{figure*}

The data reduction was carried out with \texttt{excalibuhr}\footnote{\url{https://github.com/yapenzhang/excalibuhr}} \citetext{Zhang et al., \textit{in prep.}}, a \texttt{Python} data reduction pipeline that largely follows the steps outlined in \citet{Holmberg_ea_2022}. \texttt{excalibuhr} employs similar routines as \texttt{pycrires} \citep{Stolker_ea_2023_pycrires} and which are also performed with the ESO \textit{cr2res} pipeline. The exposures were dark-subtracted, flat-fielded and the background sky emission was removed via subtraction between AB (or BA) nodding pairs. The exposures were subsequently mean-combined per nodding position (i.e. A or B). The evenly-spaced lines in the Fabry-P\'erot Etalon (FPET) observations were used to extract the slit curvature and served as a first assessment of the wavelength solution. After correcting for the slit curvature and tilt, the spectra were extracted by fitting a profile using the optimal extraction algorithm \citep{Holmberg_ea_2022}, utilising an extraction aperture of 30 pixels. The 1D spectra at the A and B nodding position were mean-combined and the blaze-function, retrieved from the flat-field exposures, was corrected for. The \texttt{excalibuhr} extraction yielded a signal-to-noise of $\sim$\,$40$ per pixel at $2.345\ \mathrm{\mu m}$. The telluric standard spectrum was extracted in a similar manner. 

A second-stage wavelength correction was carried out by fitting the telluric standard spectrum to an ESO \texttt{SkyCalc}\footnote{\url{https://www.eso.org/observing/etc/skycalc}} model \citep{Noll_ea_2012,Jones_ea_2013}. The initial wavelength solution is stretched and compressed as a 3$^\mathrm{rd}$-order polynomial via:
\begin{align}
    \lambda' = \lambda + \sum_{i=0}^{3} p_i\cdot\Big(\lambda-\langle\lambda\rangle\Big)^i, 
\end{align}
where $\lambda'$ is the new wavelength solution, $p_i$ are the polynomial coefficients, and $\langle\lambda\rangle$ is the average wavelength. The wavelength solution was found separately for each of the 7 spectral orders and 3 detectors via a $\chi^2$-minimisation, employing the Nelder-Mead simplex algorithm implemented in  \texttt{scipy.optimize.minimize} \citep{Gao_ea_2012}. The telluric standard's wavelength solution was thereafter adopted for the DENIS J0255 spectrum. As a consequence of its lower S/N and the large number of intrinsic features, the correction could not be performed on the target spectrum directly. However, the wavelength solution is not expected to change considerably as the telluric standard and target were observed in sequence.

The combined telluric transmissivity and instrumental throughput were obtained by dividing the observed, telluric standard spectrum by a PHOENIX model spectrum \citep{Husser_ea_2013}. As the high-resolution ($R=500\,000$) PHOENIX model spectra only go up to an effective temperature of $T_\mathrm{eff}=12\,000\ \mathrm{K}$, we adopted the highest-temperature spectrum and adjusted its slope to match a blackbody spectrum with the temperature of kap Eri ($T_\mathrm{eff}=14\,700\ \mathrm{K}$; \citealt{Levenhagen_ea_2006}). Any slope error introduced by this correction method is negligible, due to the weak temperature-dependency in the Rayleigh-Jeans regime. The model spectrum was Doppler-shifted to the standard star's radial velocity ($\textit{v}_\mathrm{rad}=25.5\ \mathrm{km\ s^{-1}}$; \citealt{Gontcharov_2006}) and subsequently broadened to the spectral resolution ($R=100\,000$) and rotational velocity ($\textit{v}\sin i=10\ \mathrm{km\ s^{-1}}$; \citealt{Levenhagen_ea_2006}). This procedure is performed to replicate the line wings of the Brackett $\gamma$ and $\delta$ lines present in the telluric standard's spectrum. The line cores were more poorly reproduced in the model spectrum and thus a region of $\pm1\ \mathrm{nm}$ was masked for both lines. The spectrum of DENIS J0255 was subsequently divided by the transmissivity, thereby removing telluric absorption lines and correcting for a wavelength-dependent throughput of the CRIRES$^+$ spectrograph. The deepest telluric lines are saturated and do not provide an adequate correction as a result. Hence, any pixels where the telluric transmission was lower than $\mathcal{T}<0.6$ were masked in the DENIS J0255 spectrum. A flux-calibration was performed by scaling the observed spectrum to match the 2MASS Ks-band photometry ($\mathrm{Ks}=11.56\pm0.02\ \mathrm{mag}$; \citealt{Cutri_ea_2003}) when integrated over the $\mathrm{Ks}$ filter-curve. The utilised $\mathrm{Ks}$-band magnitude is possibly discrepant from the flux at the time of measurement due to the known variability of brown dwarfs, especially at the L-T transition \citep{Wilson_ea_2014,Radigan_ea_2014,Radigan_2014}. At least, separate reductions between consecutive nodding pairs showed negligible variability throughout the hour-long observing programme. Residual differences in the observed and calibrating fluxes would be captured by a flux-scaling radius parameter in the retrieval analysis (Sect. \ref{sect:retrieval}). Finally, outliers were removed by sigma-clipping pixels beyond $>3\sigma$ from a median-filtered spectrum, using an 8-pixel wide window and the \texttt{excalibuhr}-computed flux-uncertainties as $\sigma$. Figure \ref{fig:observed_spectrum} shows the reduced spectrum of DENIS J0255 and a zoom-in of the 6th spectral order in black. The grey line displays the spectrum without correcting for the telluric absorption lines. The evenly-spaced spectral lines of $^{12}$CO are unmistakably observed in the bottom panel of Fig. \ref{fig:observed_spectrum}. The high rotational velocity of DENIS J0255 is also apparent from the line broadening, compared to the narrow telluric lines in grey. 

\subsection{Retrieval framework}\label{sect:retrieval}
For the atmospheric retrieval, we employed a Bayesian framework where the radiative transfer code \texttt{petitRADTRANS} (\texttt{pRT}; \citealt{Molliere_ea_2019b,Molliere_ea_2020,Alei_ea_2022}) is coupled with the nested sampling tool \texttt{PyMultiNest} \citep{Buchner_ea_2014} which itself is a \texttt{Python} wrapper of the \texttt{MultiNest} algorithm \citep{Feroz_ea_2009}. The computations were performed, in parallel, on the Dutch National Supercomputer Snellius\footnote{\url{https://www.surf.nl/en/dutch-national-supercomputer-snellius}}. Model emission spectra are generated by \texttt{pRT} with a number of parameters describing properties such as the thermal profile, chemical abundances, and surface gravity. We define 50 atmospheric layers between $P=10^{-6}$ and $10^{2}\ \mathrm{bar}$, equally-separated in log pressure. Collision-induced absorption from H$_2$-H$_2$ and H$_2$-He and Rayleigh scattering of H$_2$ and He are taken into account. H$_2$O, $^{12}$CO, $^{13}$CO, CH$_4$, NH$_3$, CO$_2$ and HCN are included as line opacity species. The HITEMP line lists were employed for $^{12}$CO, $^{13}$CO \citep{Li_ea_2015}, CH$_4$ \citep{Hargreaves_ea_2020}, and CO$_2$ \citep{Rothman_ea_2010}. ExoMol line lists were used for the opacity of NH$_3$ \citep{Coles_ea_2019} and HCN \citep{Harris_ea_2006,Barber_ea_2014}. For H$_2$O, both the ExoMol (POKAZATEL; \citealt{Polyansky_ea_2018}) and HITEMP \citep{Rothman_ea_2010} line lists were evaluated as is outlined in Sect. \ref{sect:H2O_line_list}. 

While \texttt{pRT} provides an implementation of physically-motivated clouds (e.g. MgSiO$_3$, Fe, etc.), we used a simple grey cloud model to not impose any assumptions on the cloud composition. Additionally, this choice was made as we did not expect to constrain any wavelength dependence of the cloud opacity over the covered wavelength range ($\Delta\lambda\sim$\,$0.5\ \mathrm{\mu m}$). We implemented the cloud model with the \texttt{give\_absorption\_opacity} functionality in \texttt{pRT}. Similar to \citet{Molliere_ea_2020}, the cloud opacity $\kappa_\mathrm{cl}$ at a pressure $P$ is computed as
\begin{align}
    \kappa_\mathrm{cl}(P) = 
    \begin{cases}
      \kappa_{\mathrm{cl},0} \left(\dfrac{P}{P_\mathrm{base}}\right)^{f_\mathrm{sed}} & P<P_\mathrm{base}, \\
      0 & P\geq P_\mathrm{base}, \\
    \end{cases}  
\end{align}
where $\kappa_{\mathrm{cl},0}$ is the opacity at the cloud base, which is set by $P_\mathrm{base}$. The cloud opacity decays above the base as a power-law controlled by $f_\mathrm{sed}$. The parameters $\kappa_{\mathrm{cl},0}$, $P_\mathrm{base}$, and $f_\mathrm{sed}$ are fit during the retrieval.

The \texttt{pRT}-generated spectra were shifted with a radial velocity $\textit{v}_\mathrm{rad}$ and subsequently broadened with a projected rotational velocity $\textit{v}\sin i$ and linear limb-darkening coefficient $\varepsilon_\mathrm{limb}$ using the \texttt{fastRotBroad} routine of \texttt{PyAstronomy}\footnote{\url{https://github.com/sczesla/PyAstronomy}} \citep{Gray_ea_2008,Czesla_ea_2019}. To speed-up the computations, the \texttt{pRT} model spectra were generated with an $\texttt{lbl\_opacity\_sampling}=3$ (i.e. $R=10^6/3$). After the rotational broadening, the spectra were down-convolved to a resolution of $R=100\,000$ to match the $0.2"$-slit observations. The observed spectrum might have an increased spectral resolution, resulting from the good seeing (e.g. \citealt{Lesjak_ea_2023}), but the retrieved rotational broadening parameters ($\textit{v}\sin i$ and $\varepsilon_\mathrm{limb}$) can largely account for any discrepancies.

\texttt{PyMultiNest} samples the parameter-space of the relevant free parameters and builds up posterior distributions from repeated likelihood evaluations between the observed and model spectra. Additionally, \texttt{PyMultiNest} computes the Bayesian evidence $\mathcal{Z}$, thus allowing for model comparisons where complexity is taken into account \citep{Feroz_ea_2009}. The constant efficiency mode was employed to allow for feasible convergence times. Following the \texttt{MultiNest} recommendations\footnote{\url{https://github.com/JohannesBuchner/MultiNest/blob/master/README}}, we used a sampling efficiency of $5\%$ in combination with the Importance Nested Sampling (INS) mode, which allows accurate evidences to be calculated \citep{Feroz_ea_2019}. The posterior distribution was sampled with 400 live points and the retrieval terminated with an evidence tolerance of $0.5$. For the fiducial model, which consists of a free-chemistry approach, the priors of the retrieved parameters are listed in Table \ref{tab:params}. In total, the fiducial retrieval fits for 32 free parameters. The following sections provide further details on the purpose of the listed parameters. 

\begin{table*}[h!]
\centering
\caption{Free parameters and the utilised prior ranges of the discussed retrievals. \label{tab:params}}
{\scriptsize
\begin{tabular}{lll|rrrrr}
\hline\hline
\textbf{Parameter} & \textbf{Description} & \textbf{Prior} & \textbf{Free-chem.} & \textbf{Quenched} & \textbf{Quenched} & \textbf{Eq.-chem.} & \textbf{HITEMP} \\
& & & \textbf{(fiducial)} & \textbf{via $\mathbf{P_{\text quench}}$} & \textbf{via $\mathbf{K_{\text zz}}$} & & \textbf{H$_{\bf2}$O} \\
\hline
$R$ [$\mathrm{R_\mathrm{Jup}}$] & Radius & $\mathcal{U}(0.4,1.5)$ & $0.78^{+0.01}_{-0.01}$ & $0.77^{+0.01}_{-0.01}$ & $0.78^{+0.01}_{-0.01}$ & $0.67^{+0.01}_{-0.01}$ & $0.85^{+0.01}_{-0.01}$ \\
$\log\ \textit{g}$ [$\mathrm{cm\ s^{-2}}$] & Surface gravity & $\mathcal{U}(4.5,6)$ & $5.27^{+0.04}_{-0.04}$ & $5.31^{+0.04}_{-0.03}$ & $5.25^{+0.04}_{-0.04}$ & $6.000^{+0.001}_{-0.003}$ & $5.74^{+0.04}_{-0.04}$ \\
$\varepsilon_\mathrm{limb}$ & Limb-darkening coefficient & $\mathcal{U}(0.2,1)$ & $0.65^{+0.03}_{-0.03}$ & $0.65^{+0.03}_{-0.03}$ & $0.66^{+0.03}_{-0.03}$ & $0.72^{+0.03}_{-0.03}$ & $0.58^{+0.03}_{-0.03}$ \\
$\textit{v}\sin i$ [$\mathrm{km\ s^{-1}}$] & Rotational velocity & $\mathcal{U}(35,50)$ & $41.05^{+0.19}_{-0.19}$ & $41.04^{+0.18}_{-0.18}$ & $41.11^{+0.18}_{-0.19}$ & $41.49^{+0.22}_{-0.21}$ & $40.45^{+0.14}_{-0.15}$ \\
$\textit{v}_\mathrm{rad}$ [$\mathrm{km\ s^{-1}}$] & Radial velocity & $\mathcal{U}(20,25)$ & $22.55^{+0.07}_{-0.07}$ & $22.56^{+0.07}_{-0.07}$ & $22.57^{+0.07}_{-0.07}$ & $22.45^{+0.09}_{-0.10}$ & $22.49^{+0.07}_{-0.07}$ \\
\hline
$\log\ \mathrm{^{12}CO}$ & VMR of $^{12}$CO & $\mathcal{U}(-10,-2)$ & $-3.32^{+0.03}_{-0.03}$ & - & & - & $-3.12^{+0.03}_{-0.03}$ \\
$\log\ \mathrm{H_2O}$ & \hspace{0.029\textwidth}"\hspace{0.029\textwidth} H$_2$O & $\mathcal{U}(-10,-2)$ & $-3.62^{+0.03}_{-0.03}$ & - & & - & $-3.36^{+0.03}_{-0.03}$ \\
$\log\ \mathrm{CH_4}$ & \hspace{0.029\textwidth}"\hspace{0.029\textwidth} CH$_4$ & $\mathcal{U}(-10,-2)$ & $-4.92^{+0.03}_{-0.03}$ & - & & - & $-4.61^{+0.03}_{-0.03}$ \\
$\log\ \mathrm{NH_3}$ & \hspace{0.029\textwidth}"\hspace{0.029\textwidth} NH$_3$ & $\mathcal{U}(-10,-2)$ & $-5.99^{+0.05}_{-0.05}$ & - & & - & $-5.75^{+0.06}_{-0.06}$ \\
$\log\ \mathrm{^{13}CO}$ & \hspace{0.029\textwidth}"\hspace{0.029\textwidth} $^{13}$CO & $\mathcal{U}(-10,-2)$ & $-5.59^{+0.11}_{-0.13}$ & - & & - & $-5.20^{+0.09}_{-0.10}$ \\
$\log\ \mathrm{CO_2}$ & \hspace{0.029\textwidth}"\hspace{0.029\textwidth} CO$_2$ & $\mathcal{U}(-10,-2)$ & $-7.99^{+1.27}_{-1.18}$ & - & & - & $-7.72^{+1.39}_{-1.36}$ \\
$\log\ \mathrm{HCN}$ & \hspace{0.029\textwidth}"\hspace{0.029\textwidth} HCN & $\mathcal{U}(-10,-2)$ & $-7.73^{+1.29}_{-1.34}$ & - & & - & $-7.11^{+1.51}_{-1.65}$ \\
$\mathrm{C/O}$ & Carbon-to-oxygen ratio & $\mathcal{U}(0,1)$ & \textcolor{darkgray}{$\mathit{0.681^{+0.005}_{-0.005}}$} & $0.631^{+0.004}_{-0.004}$ & $0.639^{+0.005}_{-0.005}$ & $0.553^{+0.004}_{-0.004}$ & \textcolor{darkgray}{$\mathit{0.656^{+0.005}_{-0.005}}$} \\
$\mathrm{[Fe/H]}$ \textcolor{darkgray}{\textit{or [C/H]}} & Metallicity & $\mathcal{U}(-1.5,1.5)$ & \textcolor{darkgray}{$\mathit{0.03^{+0.03}_{-0.03}}$} & $0.06^{+0.02}_{-0.03}$ & $0.02^{+0.03}_{-0.03}$ & $0.89^{+0.03}_{-0.03}$ & \textcolor{darkgray}{$\mathit{0.24^{+0.03}_{-0.03}}$} \\
$\log\ \mathrm{^{13}CO/^{12}CO}$ & CO isotopologue ratio & $\mathcal{U}(-10,0)$ & \textcolor{darkgray}{$\mathit{-2.27^{+0.11}_{-0.13}}$} & $-2.26^{+0.10}_{-0.12}$ & $-2.26^{+0.10}_{-0.13}$ & $-5.84^{+2.42}_{-2.49}$ & \textcolor{darkgray}{$\mathit{-2.07^{+0.08}_{-0.09}}$} \\
$\log\ P_\mathrm{quench} (\mathrm{CO,CH_4,H_2O})$ [$\mathrm{bar}$] & Quench pressure CO-CH$_4$ & $\mathcal{U}(-6,2)$ & - & $1.47^{+0.06}_{-0.06}$ & \textcolor{darkgray}{$\mathit{1.24^{+0.02}_{-0.02}}$} & - & - \\
$\log\ P_\mathrm{quench} (\mathrm{NH_3})$ [$\mathrm{bar}$] & \hspace{0.0535\textwidth}"\hspace{0.0535\textwidth} N$_2$-NH$_3$ & $\mathcal{U}(-6,2)$ & - & $0.24^{+0.11}_{-0.12}$ & \textcolor{darkgray}{$\mathit{1.29^{+0.02}_{-0.02}}$} & - & - \\
$\log\ P_\mathrm{quench} (\mathrm{HCN})$ [$\mathrm{bar}$] & \hspace{0.0535\textwidth}"\hspace{0.0535\textwidth} HCN & $\mathcal{U}(-6,2)$ & - & $-2.14^{+2.20}_{-2.28}$ & \textcolor{darkgray}{$\mathit{1.29^{+0.02}_{-0.02}}$} & - & - \\
$\log\ P_\mathrm{quench} (\mathrm{CO_2})$ [$\mathrm{bar}$] & \hspace{0.0535\textwidth}"\hspace{0.0535\textwidth} CO$_2$ & $\mathcal{U}(-6,2)$ & - & $-2.00^{+2.25}_{-2.24}$ & \textcolor{darkgray}{$\mathit{0.49^{+0.03}_{-0.03}}$} & - & - \\
$\log\ K_\mathrm{zz}$ [$\mathrm{cm^2\ s^{-1}}$] & Eddy diffusion coefficient & $\mathcal{U}(5,15)$ & - & - & $9.57^{+0.07}_{-0.08}$ & - & - \\
\hline
$\log\gamma$ & PT profile smoothing & $\mathcal{U}(-4,4)$ & $0.86^{+0.89}_{-0.73}$ & $1.17^{+0.72}_{-0.57}$ & $2.64^{+0.60}_{-0.56}$ & $1.62^{+0.76}_{-0.57}$ & $0.70^{+1.11}_{-0.90}$ \\
$\Delta\log P_\mathrm{PT}$ [$\mathrm{bar}$] & Separation of bottom knots & $\mathcal{U}(0,2)$ & $0.31^{+0.10}_{-0.09}$ & $0.23^{+0.04}_{-0.05}$ & $0.28^{+0.05}_{-0.04}$ & $0.76^{+0.02}_{-0.03}$ & $0.33^{+0.07}_{-0.09}$ \\
$T_1$ [$\mathrm{K}$] & Temperature at node $1$ & $\mathcal{U}(0,6000)$ & $3802^{+1169}_{-1135}$ & $4654^{+678}_{-783}$ & $2271^{+1447}_{-758}$ & $5451^{+378}_{-588}$ & $1975^{+1235}_{-438}$ \\
$T_2$ [$\mathrm{K}$] & \hspace{0.0535\textwidth}"\hspace{0.0535\textwidth} node $2$ & $\mathcal{U}(0,4500)$ & $2613^{+628}_{-417}$ & $3255^{+295}_{-237}$ & $4125^{+254}_{-313}$ & $1577^{+11}_{-10}$ & $1760^{+232}_{-94}$ \\
$T_3$ [$\mathrm{K}$] & \hspace{0.0535\textwidth}"\hspace{0.0535\textwidth} node $3$ & $\mathcal{U}(0,3000)$ & $1787^{+87}_{-71}$ & $1867^{+42}_{-42}$ & $2013^{+94}_{-97}$ & $1659^{+12}_{-11}$ & $1609^{+33}_{-32}$ \\
$T_4$ [$\mathrm{K}$] & \hspace{0.0535\textwidth}"\hspace{0.0535\textwidth} node $4$ & $\mathcal{U}(0,2000)$ & $1507^{+41}_{-47}$ & $1545^{+22}_{-22}$ & $1529^{+15}_{-17}$ & $1350^{+13}_{-12}$ & $1358^{+44}_{-36}$ \\
$T_5$ [$\mathrm{K}$] & \hspace{0.0535\textwidth}"\hspace{0.0535\textwidth} node $5$ & $\mathcal{U}(0,2000)$ & $1107^{+43}_{-50}$ & $1149^{+25}_{-25}$ & $1118^{+21}_{-29}$ & $1145^{+6}_{-6}$ & $1039^{+30}_{-23}$ \\
$T_6$ [$\mathrm{K}$] & \hspace{0.0535\textwidth}"\hspace{0.0535\textwidth} node $6$ & $\mathcal{U}(0,2000)$ & $729^{+32}_{-33}$ & $743^{+23}_{-23}$ & $777^{+24}_{-25}$ & $979^{+8}_{-8}$ & $805^{+27}_{-32}$ \\
$T_7$ [$\mathrm{K}$] & \hspace{0.0535\textwidth}"\hspace{0.0535\textwidth} node $7$ & $\mathcal{U}(0,2000)$ & $1077^{+539}_{-565}$ & $1131^{+522}_{-568}$ & $1019^{+570}_{-603}$ & $1427^{+378}_{-430}$ & $995^{+538}_{-569}$ \\
\hline
$\log\ \kappa_{\mathrm{cl},0}$ [$\mathrm{cm^2\ g^{-1}}$] & Opacity at cloud base & $\mathcal{U}(-10,3)$ & $-4.81^{+3.19}_{-3.14}$ & $-4.77^{+3.00}_{-2.97}$ & $-4.52^{+3.08}_{-3.26}$ & $-4.48^{+3.57}_{-3.62}$ & $-4.59^{+3.12}_{-3.08}$ \\
$\log\ P_\mathrm{base}$ [$\mathrm{bar}$] & Cloud base pressure & $\mathcal{U}(-6,3)$ & $-1.69^{+2.93}_{-2.53}$ & $-1.79^{+2.82}_{-2.47}$ & $-2.05^{+3.02}_{-2.48}$ & $-1.64^{+3.00}_{-2.72}$ & $-1.87^{+2.86}_{-2.39}$ \\
$f_{\mathrm{sed}}$ & Cloud decay power & $\mathcal{U}(0,20)$ & $10.06^{+5.65}_{-5.76}$ & $9.76^{+5.68}_{-5.42}$ & $10.11^{+5.97}_{-5.88}$ & $10.19^{+6.27}_{-6.24}$ & $10.28^{+5.71}_{-6.02}$ \\
\hline
$a_1$ & GP amplitude for order $1$ & $\mathcal{U}(0.1,0.8)$ & $0.62^{+0.04}_{-0.04}$ & $0.60^{+0.04}_{-0.04}$ & $0.62^{+0.05}_{-0.04}$ & $0.69^{+0.04}_{-0.04}$ & $0.79^{+0.01}_{-0.01}$ \\
$a_2$ & \hspace{0.0625\textwidth}"\hspace{0.0625\textwidth} order $2$ & $\mathcal{U}(0.1,0.8)$ & $0.45^{+0.03}_{-0.03}$ & $0.44^{+0.03}_{-0.03}$ & $0.44^{+0.03}_{-0.03}$ & $0.65^{+0.04}_{-0.04}$ & $0.79^{+0.01}_{-0.01}$ \\
$a_3$ & \hspace{0.0625\textwidth}"\hspace{0.0625\textwidth} order $3$ & $\mathcal{U}(0.1,0.8)$ & $0.32^{+0.02}_{-0.02}$ & $0.32^{+0.02}_{-0.02}$ & $0.32^{+0.02}_{-0.02}$ & $0.38^{+0.03}_{-0.02}$ & $0.44^{+0.02}_{-0.02}$ \\
$a_4$ & \hspace{0.0625\textwidth}"\hspace{0.0625\textwidth} order $4$ & $\mathcal{U}(0.1,0.8)$ & $0.30^{+0.02}_{-0.02}$ & $0.31^{+0.02}_{-0.02}$ & $0.32^{+0.02}_{-0.02}$ & $0.38^{+0.03}_{-0.03}$ & $0.31^{+0.02}_{-0.02}$ \\
$a_5$ & \hspace{0.0625\textwidth}"\hspace{0.0625\textwidth} order $5$ & $\mathcal{U}(0.1,0.8)$ & $0.32^{+0.02}_{-0.02}$ & $0.32^{+0.02}_{-0.02}$ & $0.33^{+0.02}_{-0.02}$ & $0.45^{+0.03}_{-0.03}$ & $0.36^{+0.02}_{-0.02}$ \\
$a_6$ & \hspace{0.0625\textwidth}"\hspace{0.0625\textwidth} order $6$ & $\mathcal{U}(0.1,0.8)$ & $0.50^{+0.03}_{-0.02}$ & $0.51^{+0.03}_{-0.02}$ & $0.51^{+0.03}_{-0.03}$ & $0.795^{+0.004}_{-0.007}$ & $0.51^{+0.03}_{-0.03}$ \\
$a_7$ & \hspace{0.0625\textwidth}"\hspace{0.0625\textwidth} order $7$ & $\mathcal{U}(0.1,0.8)$ & $0.65^{+0.03}_{-0.03}$ & $0.65^{+0.03}_{-0.03}$ & $0.67^{+0.03}_{-0.03}$ & $0.794^{+0.005}_{-0.008}$ & $0.795^{+0.004}_{-0.006}$ \\
$\ell$ [$\mathrm{km\ s^{-1}}$] & GP length-scale & $\mathcal{U}(10,40)$ & $23.87^{+1.04}_{-0.96}$ & $23.44^{+0.97}_{-0.94}$ & $23.60^{+1.02}_{-0.93}$ & $23.94^{+0.65}_{-0.68}$ & $26.90^{+0.82}_{-0.75}$ \\
\hline\hline
 & & $\mathbf{ln\ B_m}$ & - & $-4.2\ (-3.4\sigma)$ & $-12.1\ (-5.3\sigma)$ & $-416\ (-29.0\sigma)$ & $-315\ (-25.2\sigma)$ \\
\end{tabular}
}
\tablefoot{The five rightmost columns indicate the retrieved parameters for the free-chemistry, quenched equilibrium (via $P_\mathrm{quench}$ and $K_\mathrm{zz}$), and un-quenched equilibrium approaches discussed in Sect. \ref{sect:diseq_chem}, as well as the retrieval carried out with the HITEMP H$_2$O line list outlined in \ref{sect:H2O_line_list}. Grey cursive values were not retrieved as free parameters, but derived from the given results. The Bayes factors ($\ln{B_m}$) on the bottom row were calculated with respect to the fiducial model.}
\end{table*}

\subsubsection{Likelihood \& covariance}
We adopt a likelihood formalism similar to \citet{Gibson_ea_2020}. For each order-detector pair, the log-likelihood is calculated as:
\begin{align}
    \ln\mathcal{L} &= -\frac{1}{2}\left(N\ln(2\pi) + \ln(|\vec{\Sigma}_0|) + N\ln(s^2) + \frac{1}{s^2}\vec{R}^T \vec{\Sigma}_0^{-1} \vec{R}\right), \label{eq:ln_L}
\end{align}
with $N$ the number of pixels, $\vec{\Sigma}_0$ the covariance matrix, comprising of the flux-uncertainty per pixel and the covariance between the pixels. $\vec{R}$ denotes the residuals between the observed and model spectrum. Following the linear scaling implementation of \citet{Ruffio_ea_2019}, the residuals are computed with:
\begin{align}
    \vec{R} &= \vec{d} - \phi \vec{m}, 
\end{align}
where $\vec{d}$ and $\vec{m}$ are vectors of the observed and model spectra, respectively. The flux-scaling parameter $\phi$ is optimised for each order and detector except for the first, whose flux is scaled by the free radius parameter $R$. Hence, the scaling is performed relative to the first order-detector pair. The rationale behind applying this separate flux-scaling is that it corrects for small, intra-order errors in the slope of the spectrum, possibly introduced by the gain- or blaze-correction. The optimal $\tilde{\phi}$ is found by solving
\begin{align}
    \vec{m}^T \vec{\Sigma}_0^{-1} \vec{m}\cdot \tilde{\phi} = \vec{m}^T \vec{\Sigma}_0^{-1} \vec{d}. \label{eq:phi}
\end{align}
In Eq. \ref{eq:ln_L}, $s^2$ is a covariance scaling parameter that is optimised for each order-detector pair. For the optimally-scaled residuals, $\tilde{s}^2$ is found via
\begin{align}
    \tilde{s}^2 &= \frac{1}{N} \vec{R}^T \vec{\Sigma}_0^{-1} \vec{R} \bigg|_{\phi=\tilde{\phi}}. \label{eq:s}
\end{align}
Effectively, the total covariance matrix $\vec{\Sigma}=s^2\vec{\Sigma}_0$ is scaled so that the reduced chi-squared is equal to 
$\chi_\mathrm{red}^2=\vec{R}^T\vec{\Sigma}^{-1}\vec{R}/N=1$. As a result, $s$ provides an assessment of the over- or under-estimation of the flux-uncertainties under the assumption of a perfect model fit and when only considering uncorrelated noise. Appendix D of \citet{Ruffio_ea_2019} provides more details concerning the method by which the optimal $\tilde{\phi}$ and $\tilde{s}^2$ are found. Both $\phi$ and $s^2$ are optimised for each model during the retrieval.

In the case where the pixel-measurements are uncorrelated, the covariance matrix would consist only of diagonal, squared uncertainty terms (i.e. $\Sigma_{0,ij}=\delta_{ij}\sigma_i^2$). However, covariance can arise from multiple sources \citep{Czekala_ea_2015,Zhang_Z_ea_2021,Iyer_ea_2023}, including oversampling of the instrumental line spread function by several pixels ($\sim$\,$3$ pixels for CRIRES$^+$), rotational broadening, systematics, or through imperfections in the model spectrum (e.g. uncertain opacities). Therefore, any deficiencies in a modelled spectral feature results in correlated residuals over multiple adjacent pixels. The retrieved parameters can be biased and their uncertainties will be underestimated when only considering uncorrelated noise. This problem can be addressed by Gaussian Processes (GPs), allowing for the modelling of off-diagonal covariance matrix-elements within the Bayesian framework \citep{Czekala_ea_2015,Kawahara_ea_2022}. In this work, we employed a global radial basis function (RBF) kernel, similar to \citet{Kawahara_ea_2022}. The covariance of pixels $i$ and $j$ can then be described as
\begin{align}
    \Sigma_{0,ij} &= \underbrace{\delta_{ij}\sigma_i^2}_{\substack{\text{un-correlated}\\\text{component}}} + \underbrace{a^2 \sigma_{\mathrm{eff},ij}^2 \exp{\left(-\dfrac{r_{ij}^2}{2\ell^2}\right)}}_\text{correlated component (GP)}, \label{eq:Sigma_0ij}
\end{align}
where $\delta_{ij}$ is the Kronecker delta, $\sigma_i$ the flux-uncertainty of pixel $i$. Respectively, $a$ and $\ell$ are the GP amplitude and length-scale, which are sampled as free parameters during the retrieval. The pixel separation $r_{ij}$ used in this study is given in velocity-space:
\begin{align}
    r_{ij} &= 2c \left|\frac{\lambda_i - \lambda_j}{\lambda_i + \lambda_j}\right|, 
\end{align}
with $c$ as the speed of light. The effective uncertainty $\sigma_{\mathrm{eff},ij}$ used in this work is given by the arithmetic mean of the variances of pixels $i$ and $j$:
\begin{align}
    \sigma_{\mathrm{eff},ij}^2 &= \frac{1}{2}\left(\sigma_i^2+\sigma_j^2\right). 
\end{align}
Thus, the amplitude $a$ scales the off-diagonal covariance terms relative to their respective diagonal elements. The GP amplitude $a$ is retrieved separately for each order (i.e. $a_1$, $a_2$, etc.), but to minimise the number of free parameters, and since we do not expect the correlation length to differ significantly, we only retrieved one length-scale $\ell$. Correlation between detectors (and orders) was not considered due to the nanometer-wide gaps. We emphasise that the uncertainty scaling parameter $s^2$ needs to be multiplied with the full $\vec{\Sigma}_0$ (in Eq. \ref{eq:ln_L}) to be analytically marginalised over, and thus applies to both components of Eq. \ref{eq:Sigma_0ij}. 

Banded Cholesky decompositions are applied to the covariance matrices to compute Eq. \ref{eq:ln_L}. The linear algebra of equations \ref{eq:ln_L}, \ref{eq:phi} and \ref{eq:s} would be computationally expensive because the covariance matrices $\vec{\Sigma}_0$ have considerable sizes per order-detector pair ($2048\times2048$ without masking). By the nature of the correlation, the largest values are concentrated near the diagonal. For that reason, only the elements with $r_{ij}\leq 5\ell$ were considered in the banded Cholesky decomposition, thereby reducing the number of computations to be made. Appendix \ref{app:GPs} presents the results of a retrieval where the covariance is not modelled and this results in considerable biases to the retrieved parameters.

\subsubsection{Chemistry} \label{sect:chemistry}
In this work, we considered multiple approaches to modelling the chemical abundances, or volume-mixing ratios (VMRs). 
\begin{enumerate}[-]
    \item \textit{Free-chemistry}: The fiducial retrieval consists of a free-chemistry approach, where the abundances of relevant chemical species are allowed to vary in order to find the best fit to the data without making assumptions about the thermo-chemical state of the atmosphere. The molecular abundances are kept vertically constant to reduce the number of free parameters. As part of this approach, the He abundance is held constant at $\mathrm{VMR_{He}}=0.15$ and the abundance of H$_2$ adjusts to obtain a total VMR$_\mathrm{tot}$ equal to unity. 
    \item \textit{Un-quenched equilibrium chemistry}: Aside from free-chemistry, the chemical equilibrium implementation of \texttt{pRT} \citep{Molliere_ea_2017} was also tested. With this method, the individual chemical abundances are obtained by interpolating a pre-computed chemical equilibrium table which depends on the pressure, temperature, metallicity ($\mathrm{[Fe/H]}$), and carbon-to-oxygen ratio ($\mathrm{C/O}$). The metallicity and $\mathrm{C/O}$-ratio are retrieved as free parameters in this set-up. In contrast to the free-chemistry approach, the chemical equilibrium abundances are not constant with altitude. 
    \item \textit{Quenched equilibrium chemistry ($P_\mathrm{quench}$)}: Additionally, we tested the effect of a simple quenching implementation, where we started with the same equilibrium chemistry retrieval as before, but certain molecular abundances are held constant above retrieved quenching pressures $P_\mathrm{quench}$. Hence, the upper atmosphere is driven out of chemical equilibrium. This assumption of disequilibrium at low pressures can be made since the vertical mixing timescale could become shorter than the chemical reaction timescales of the CO-CH$_4$ and N$_2$-NH$_3$ networks, for example \citep{Visscher_ea_2011,Zahnle_ea_2014}. Following \citet{Zahnle_ea_2014}, we implement four quenching pressures as retrieved parameters: $P_\mathrm{quench}(\mathrm{CO,CH_4,H_2O})$, $P_\mathrm{quench}(\mathrm{NH_3})$, $P_\mathrm{quench}(\mathrm{HCN})$, and $P_\mathrm{quench}(\mathrm{CO_2})$, where the involved molecules are denoted in parentheses.
    \item \textit{Quenched equilibrium chemistry ($K_\mathrm{zz}$)}: Lastly, \citet{Zahnle_ea_2014} provide prescriptions for the chemical timescales $\tau_\mathrm{chem}(P,T,\mathrm{[Fe/H]})$ of the four aforementioned reaction networks. This allows us to evaluate the pressures at which the chemical- and mixing timescales are equal, and thus where the quench points are located. For each model, the mixing timescale is calculated as
    \begin{align}
        \tau_\mathrm{mix} = \frac{L^2}{K_\mathrm{zz}} = \frac{(\alpha H)^2}{K_\mathrm{zz}} = \frac{\alpha^2}{K_\mathrm{zz}}\left(\frac{k_\mathrm{B}T}{\mu m_\mathrm{p} \textit{g}}\right)^2, \label{eq:tau_mix}
    \end{align}
    where $K_\mathrm{zz}$, the vertical eddy diffusion coefficient, is a free parameter in this fourth retrieval. The length scale $L=\alpha H$ is defined as the product of the scale height $H$ and a factor $\alpha$, allowing mixing lengths to be shorter than the scale height, as is found by \citet{Smith_1998} and \citet{Ackerman_ea_2001}. We adopt $\alpha=1$, so any over-estimation of the length scale $L$ will translate into the retrieved $K_\mathrm{zz}$. In comparison to the previous model, this $K_\mathrm{zz}$-retrieval requires a single parameter, rather than four, to describe the quenching.
\end{enumerate}
The differences in the results for the free-chemistry, un-quenched, and quenched equilibrium models are discussed in Sect. \ref{sect:diseq_chem}.

\subsubsection{Pressure-temperature profile} \label{sect:PT}
We present an updated parameterisation of the pressure-temperature (PT) profile, akin to that of \citet{Line_ea_2015}. A number of knots are defined in pressure space at which the temperatures are retrieved as free parameters. A cubic spline interpolation is subsequently used to determine the temperatures at each of the modelled layers. Depending on the number and spacing of the fitted knots, such a parameterisation can result in unphysical oscillations (e.g. \citealt{Rowland_ea_2023}). A smoothing can be applied to remedy this problem. The \citet{Line_ea_2015} parameterisation avoids setting an \textit{a priori} degree of smoothness by implementing a log-likelihood penalty based on the discrete second derivative of the temperature knots. The penalty is weighted by an additional free parameter $\gamma$, which tunes the smoothing and can allow for more oscillations if the data warrants it. 

We first tested the \citet{Line_ea_2015} parameterisation, but we encountered some issues. First, the definition of the log-likelihood penalty requires the knots to be spaced equally in log-pressure. \citet{Line_ea_2015} choose 15 knots between $P=10^{-3}$ and $315\ \mathrm{bar}$ to resolve the temperature gradient in the photosphere, but such a large number of free parameters (combined with the parameters for chemistry, cloud structure, etc.) affects the convergence time of \texttt{MultiNest}. In addition, a considerable number of free parameters are used to describe the temperature at pressures above and below the photosphere, which can result in overfitting. Secondly, the log-likelihood penalty is applied to the discrete second derivative in $T$-$\log(P)$ space. In the absence of information, the PT profile will therefore favour a linear solution in $T$-$\log(P)$. However, at low altitudes this is not representative of the convective adiabat of a self-consistent PT profile, which is linear in $\log(T)$-$\log(P)$ space. As a result, the PT profile can become too isothermal compared to the convective region of a self-consistent model (e.g. \citealt{Line_ea_2015,Line_ea_2017}). Applying a penalty to the second derivative in $\log(T)$-$\log(P)$ space could offer a solution to this problem, but this can result in temperature gradients which are too steep in the upper atmosphere and photosphere (e.g. Fig. 3 of \citealt{Rowland_ea_2023}). Hence, we decided to penalise the discrete third derivative in $\log(T)$-$\log(P)$.

\begin{figure*}[h!]
    \centering
    \includegraphics[width=17cm]{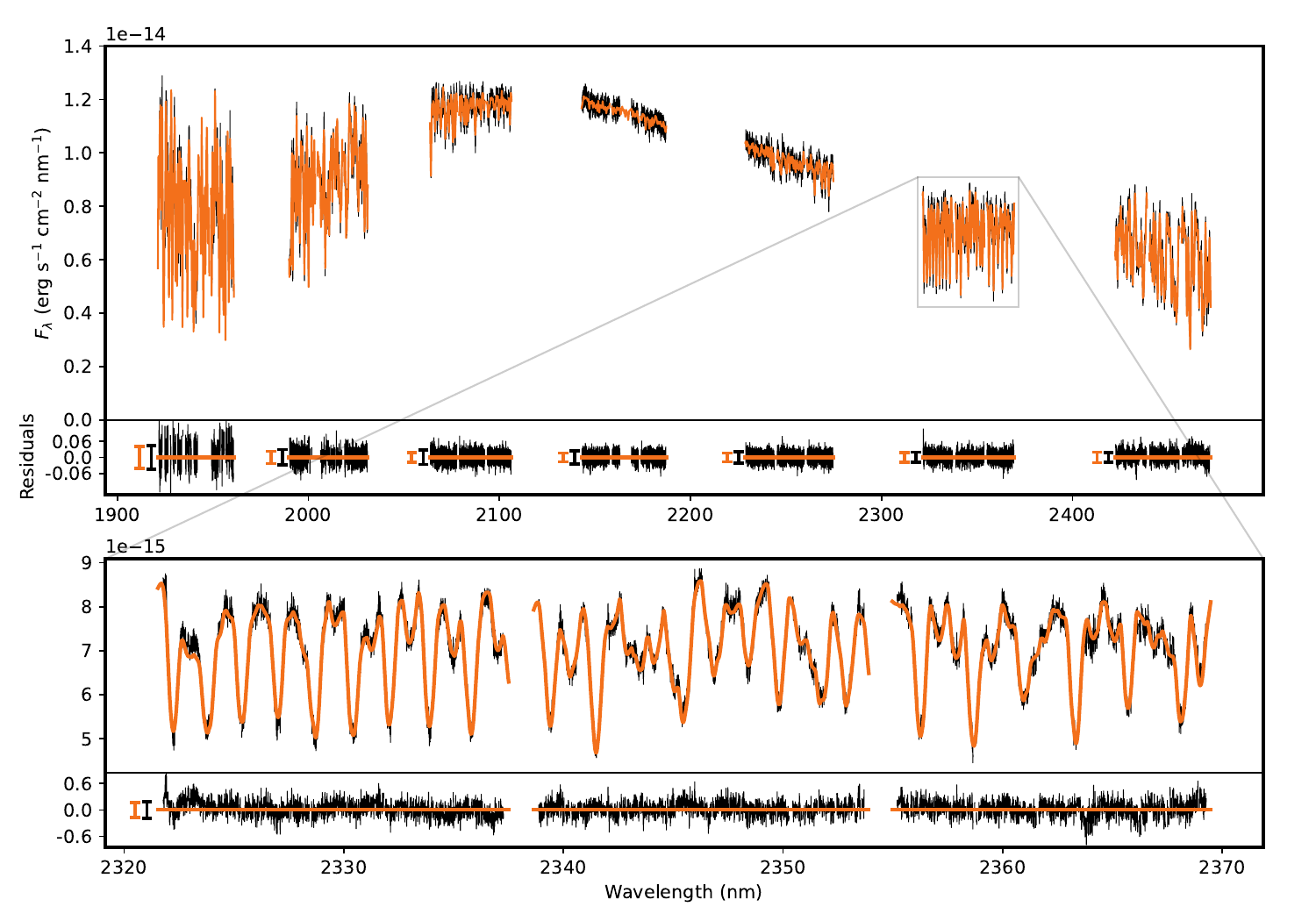}
    \caption{Best-fitting spectrum of the fiducial model, compared to the observed spectrum of DENIS J0255. \textit{Upper panels}: similar to Fig. \ref{fig:observed_spectrum}, the observed spectrum is plotted in black. The orange line displays the model spectrum and the second panel shows the residuals between the observed and model spectra in black. The mean photon noise of each order is indicated with a black error bar. The orange error bar shows the modelled uncertainty, defined as $\big\langle\sqrt{\mathrm{diag}(s^2\vec{\Sigma}_0)}\big\rangle$ per order. \textit{Lower panels}: zoom-in of the 6th spectral order, showing several correlated residual structures in the bottom panel.}
    \label{fig:bestfit_spectrum}
\end{figure*}

The general P-splines formalism of \citet{Li_ea_2022} is employed to compute the log-likelihood penalty:
\begin{align}
    \ln\mathcal{L}_\mathrm{penalty} &= -\frac{\mathrm{PEN}^{(3)}_\mathrm{gps}}{2\gamma}  - \frac{1}{2}\ln(2\pi\gamma) = -\frac{1}{2\gamma} \|\vec{D}_3\vec{C}\|^2 - \frac{1}{2}\ln(2\pi\gamma), \label{eq:ln_L_pen}
\end{align}
where $\vec{D}_3$ is the third-order general difference matrix, which computes the variations between the temperature knots and weighs them based on their separation in log pressure. Thus, a difference between $\log(T_i)$ and $\log(T_{i+1})$ results in a larger penalty if the node separation is small. Consequently, the knots can be non-uniformly distributed in pressure space which enables a higher density of knots near the photosphere. In fact, the pressure knots are not required to remain stationary, as long as $\vec{D}_3$ is re-calculated for each change of node separation. Allowing the knots to vary in temperature as well as pressure can help to avoid \textit{a priori} assumptions of the location of the photosphere. For that reason, we define the pressure knots as 
\begin{align}
    \begin{pmatrix}
        \log P_1 \\
        \log P_2 \\
        \vdots \\
        \log P_7 \\
    \end{pmatrix} &= \begin{pmatrix}
        2 \\
        2 \\
        1.5 \\
        1 \\
        0.25 \\
        -0.75 \\
        -6 \\
    \end{pmatrix} - 
    \Delta\log P_\mathrm{PT}\begin{pmatrix}
        0 \\
        1 \\ 
        1 \\
        \vdots \\
        1 \\
        0 \\
    \end{pmatrix}, 
\end{align}
where $\Delta\log P_\mathrm{PT}$ is the separation between the first and second knots at the bottom of the modelled atmosphere. This value is retrieved as a free parameter (see Table \ref{tab:params}) and thus permits the interior knots to shift vertically, towards the photosphere, where the highest resolution is required. To minimise the number of free parameters, the separation between adjacent interior knots is kept constant with an increasing separation as the knots probe the upper atmosphere. In Eq. \ref{eq:ln_L_pen}, $\vec{C}$ is the vector of coefficients that describe the cubic spline interpolation. The method by which $\vec{D}_3$ and $\vec{C}$ are computed is outlined in Appendix \ref{app:PT_def} and \citet{Li_ea_2022}. In Eq. \ref{eq:ln_L_pen}, $\gamma$ is a retrieved, free parameter that scales the contribution of the log-likelihood penalty when it is added to Eq. \ref{eq:ln_L}. Contrary to the inverse gamma prior of \citet{Line_ea_2015}, we imposed a log-uniform prior on $\gamma$ (see Table \ref{tab:params}) because this will similarly favour small values which smooth the PT profile. 

The updated PT profile parameterisation was tested on a synthetic spectrum, generated with a Sonora Bobcat PT profile \citep{Marley_ea_2021}. The configuration and results of this retrieval are outlined in Appendix \ref{app:PT_synth}. In summary, this PT parameterisation manages to retrieve the input parameters and thermal profile. Hence, we expect to be able to constrain the atmospheric properties of DENIS J0255 from its CRIRES$^+$ spectrum. 

\section{Results}\label{sect:results}
\begin{figure*}[h!]
    \centering
    \includegraphics[width=17cm]{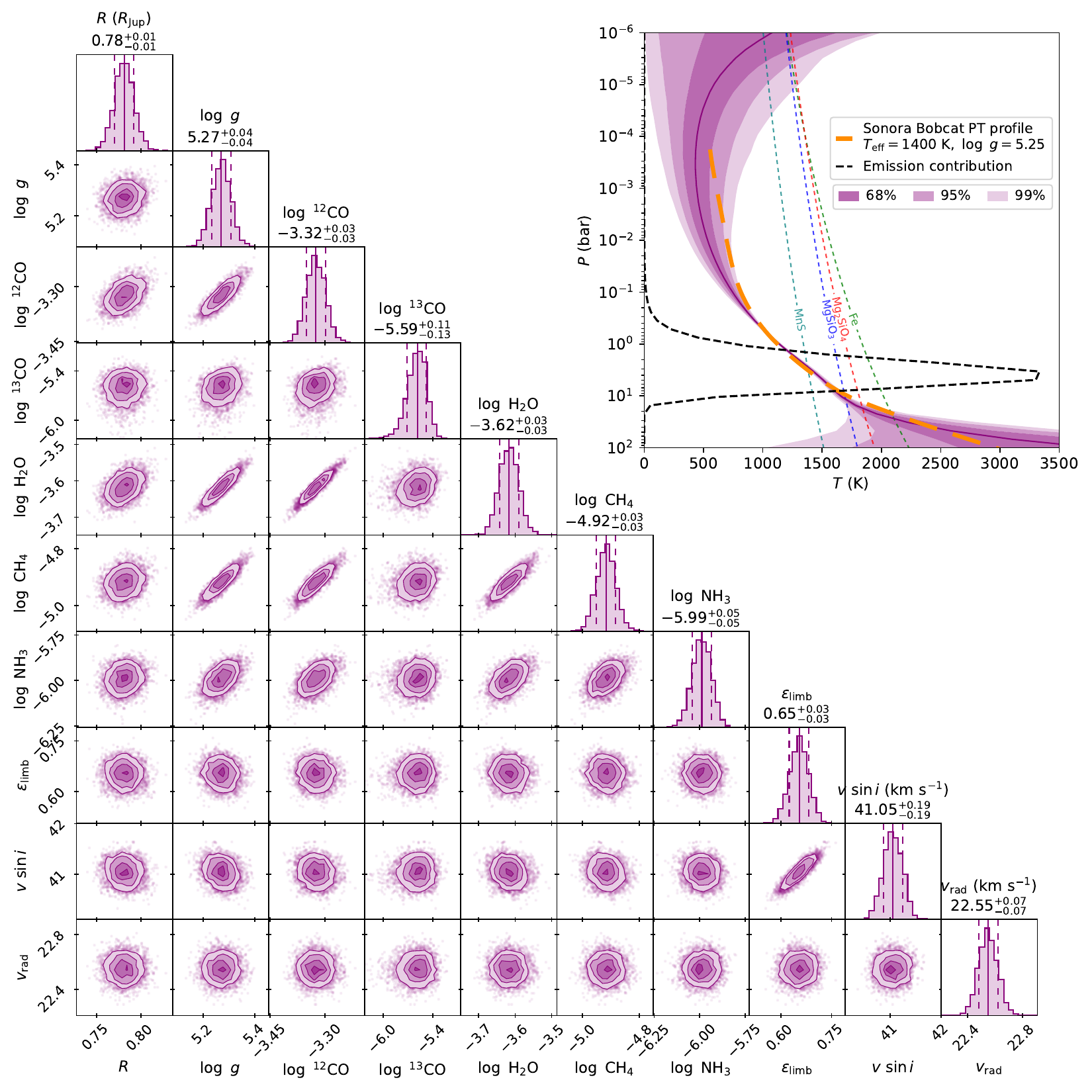}
    \caption{Posterior distributions and thermal structure of DENIS J0255's atmosphere, retrieved using the fiducial model. \textit{Upper right panel}: retrieved confidence envelopes of the PT profile. The solid purple line shows the median profile, with the shaded regions showing the 68, 95 and 99.7\% confidence intervals. For comparison, a Sonora Bobcat PT profile is shown with an orange line, in addition to condensation curves of several cloud species (MnS: \citealt{Visscher_ea_2006}; MgSiO$_3$, Mg$_2$SiO$_4$, Fe: \citealt{Visscher_ea_2010}). \textit{Lower left panels}: retrieved physical, chemical and kinematic parameters and their correlation. The posterior distributions of unconstrained parameters, in addition to those describing the Gaussian Processes or the PT profile, are omitted for clarity. The vertical dashed lines in the histograms indicate the 16th and 84th percentile, while the solid line shows the median value.}
    \label{fig:fiducial_corner}
\end{figure*}

Figure \ref{fig:bestfit_spectrum} shows the best-fitting spectrum resulting from the fiducial, free-chemistry retrieval. This fiducial model provides an excellent fit to the data, as is visible from the residuals and the zoom-in of the 6th spectral order. While the residuals are near the level expected from the S/N of the spectrum, Fig. \ref{fig:bestfit_spectrum} does show some low-order deviations, thereby highlighting the importance of accounting for pixel-to-pixel correlation using Gaussian Processes. The error bars in Fig. \ref{fig:bestfit_spectrum} show that the diagonal of the model covariance matrices $\vec{\Sigma}$ are somewhat reduced compared to the photon noise, except for the reddest order. This indicates an over-estimation of the photon noise by a factor $\sim$\,$1.3$. On the other hand, the off-diagonal elements are increased in the modelled covariance. The posterior distributions of constrained parameters and the retrieved thermal profile are shown in Figure \ref{fig:fiducial_corner}. Additionally, the fourth column of Table \ref{tab:params} lists the retrieved values. It should be noted that the retrieved posteriors are likely too narrow, as this is an established issue for \texttt{MultiNest} retrievals \citep{Buchner_2016,Ardevol_Martinez_ea_2022,Vasist_ea_2023,Latouf_ea_2023}. Our choice for a low sampling efficiency ($5\%$) and constant efficiency mode could contribute to the under-dispersion \citep{Chubb_ea_2022}, but this was necessary to achieve reasonable convergence times.

\subsection{Bulk properties, temperature \& clouds}
Interestingly, within the order-detector pairs, the slope of the K-band spectrum \citep{Canty_ea_2013} provides sufficient information to constrain the surface gravity at $\log\textit{g}\sim$\,$5.3$, in line with the high surface gravities ($\log\textit{g}\gtrsim$\,$5$) found by other studies at lower resolving powers \citep{Cushing_ea_2008,Tremblin_ea_2016,Charnay_ea_2018,Lueber_ea_2022}. However, there does exist a weak anti-correlation between $\log\textit{g}$ and $\textit{v}\sin\textit{i}$ (Pearson correlation: $r=-0.20$) due to a shared line-broadening effect. The retrieved radius of $R\sim$\,$0.8\ R_\mathrm{Jup}$ is consistent with earlier model comparisons as well \citep{Tremblin_ea_2016,Charnay_ea_2018}, but we note that this is effectively a retrieval of the 2MASS photometry owing to the basic flux calibration (Sect. \ref{sect:obs_and_reduction}). The rotational- and radial velocity measurements also agree with previous work \citep{Basri_ea_2000,Mohanty_ea_2003,Zapatero_Osorio_ea_2006}, whereas the retrieved $\varepsilon_\mathrm{limb}$ provides the first assessment of limb-darkening for this brown dwarf. As expected, the confidence envelopes of the PT profile show tight constraints near the photosphere (0.1 - 10 bar), indicated by the emission contribution function, and large uncertainties at the higher and lower altitudes which are not probed by the K-band spectrum. The Sonora Bobcat PT profile (orange line; \citealt{Marley_ea_2021}) shown for comparison in Fig. \ref{fig:fiducial_corner} is generally located within the retrieved confidence envelope, signalling an overall agreement with this self-consistent model. The high rotational broadening of DENIS J0255 means that the line cores are shallower and thus do not provide much information about higher altitudes. Hence, on top of the high surface gravity, the emission contribution function is compressed more than it would be for a slowly-rotating object. The fiducial model fails to find a solution with a significant grey cloud opacity as $\kappa_\mathrm{cl}\gtrsim$\,$ 1\ \mathrm{cm^2g^{-1}}$ is excluded for a cloud base near the photosphere. The validity of this cloud-free solution is discussed in more detail in Sect. \ref{sect:cloud_effects}.

\subsection{Detection of molecules}
The fiducial retrieval finds constraints on the abundances of H$_2$O, $^{12}$CO, CH$_4$, NH$_3$ and $^{13}$CO, as presented in Fig. \ref{fig:fiducial_corner}. Upper limits are found for the volume-mixing ratios of CO$_2$ and HCN at $10^{-6.0}$ and $10^{-5.6}$ ($97.5$-th percentile), respectively. The retrieved abundances show a strong correlation between species, particularly for H$_2$O, $^{12}$CO and CH$_4$, as is commonly seen for high-resolution spectroscopy (e.g. \citealt{Gibson_ea_2022,Gandhi_ea_2023a}). To determine whether the constraints on CH$_4$, NH$_3$ and $^{13}$CO constitute detections, we carry out three additional retrievals with the same set-up as the fiducial model, but we exclude one species at a time. By comparing the Bayesian evidence ($\mathcal{Z}$) with that of the fiducial model, we determine to what extent the inclusion of a molecule is favoured. The logarithm of the Bayes factor ($B_m$), or the difference in log-evidence
\begin{align}
    \ln{B_m} &= \ln{\mathcal{Z}_\mathrm{fiducial}} - \ln{\mathcal{Z}_{\mathrm{w/o\ species}\ X_i}}, 
\end{align}
is translated into a detection significance following \citet{Benneke_ea_2013}. We find that the inclusion of CH$_4$, NH$_3$ and $^{13}$CO is favoured at a significance of $23.1\sigma$, $7.9\sigma$ and $2.7\sigma$, respectively. 

\begin{figure}[h!]
    \resizebox{\hsize}{!}{\includegraphics[width=17cm]{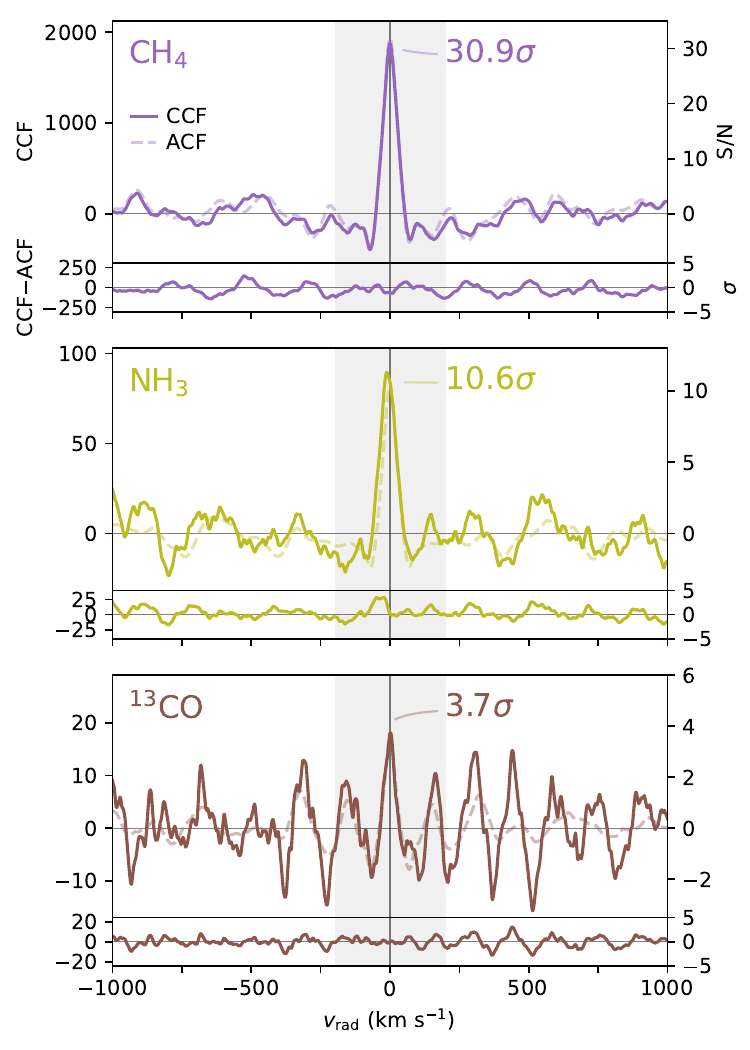}}
    \caption{Cross-correlation analysis of CH$_4$, NH$_3$, and $^{13}$CO. For each species, the upper panel shows the CCF (solid) and ACF (dashed) with the signal-to-noise indicated on the right-hand side. The lower panels present the residuals of CCF$-$ACF which are used to estimate the noise outside of the expected peak. The cross-correlation was performed on the observed spectrum centred in its rest frame, thus resulting in peaks around $0\ \mathrm{km\ s^{-1}}$.}
    \label{fig:CCF}
\end{figure}

We also carry out a cross-correlation analysis as a further robustness test of the detections of CH$_4$, NH$_3$ and $^{13}$CO. Similar to \citet{Zhang_ea_2021b}, the cross-correlation function is computed between the data residuals, that is the observed spectrum minus the fiducial model without the opacity from a certain species, and the template for that species. The species' template is calculated by taking the difference between the complete, fiducial model spectrum and the fiducial model without that molecule's opacity, thus resulting in the contribution relative to the complete model. The fiducial retrieval is used, rather than the species-excluding retrievals described above, as the latter are likely compensating for the missing opacity by changing the abundance of other species which would result in a spurious cross-correlation signal. The data residuals and species templates are high-pass filtered, using a Gaussian filter with $\sigma=300\ \mathrm{pixels}$, to remove broad features. At each radial velocity shift, and for each order-detector pair, we compute the cross-correlation coefficient as: 
\begin{align}
    CC &= \frac{1}{\tilde{s}^2} \vec{M}^T \vec{\Sigma}_0^{-1} \vec{R}, 
\end{align}
where $\vec{M}$ and $\vec{R}$ are the high-pass filtered species' contribution and data residuals, respectively. The cross-correlation is weighted by the non-diagonal covariance matrix $\vec{\Sigma}_0$, thereby placing more importance on separated features. The cross-correlation coefficients of all order-detector pairs are integrated per radial velocity shift. Figure \ref{fig:CCF} depicts the cross-correlation functions (CCFs) for the three considered molecules, in addition to the auto-correlation function (ACF) of the model template. The cross-correlation coefficients are calculated between $\pm1000\ \mathrm{km\ s^{-1}}$ in steps of $1\ \mathrm{km\ s^{-1}}$, in DENIS J0255's rest frame. In an attempt to limit the effect of auto-correlation, we compute the standard deviation on the residuals of $\mathrm{CCF}-\mathrm{ACF}$, excluding the values within $-200<\textit{v}_\mathrm{rad}<200\ \mathrm{km\ s^{-1}}$. The obtained noise estimate is subsequently used to compute the cross-correlation signal-to-noise which is shown on the right-hand side of Fig. \ref{fig:CCF}. Notably, the CCF of NH$_3$ peaks somewhat to the left of $0\ \mathrm{km\ s^{-1}}$, namely at $-12\ \mathrm{km\ s^{-1}}$, which might be the result of modelled lines being offset from their true wavelengths. The periodic line structure of $^{13}$CO produces the additional lobes observed in its CCF and ACF. The cross-correlation detection significances are established at $30.9\sigma$, $10.6\sigma$, and $3.7\sigma$ for CH$_4$, NH$_3$, and $^{13}$CO, respectively. From the Bayesian model selection and cross-correlation analyses, we conclude that the presence of CH$_4$ and NH$_3$ is robustly identified, while there is tentative evidence for $^{13}$CO.

\begin{figure*}[h!]
    \centering
    \includegraphics[width=17cm]{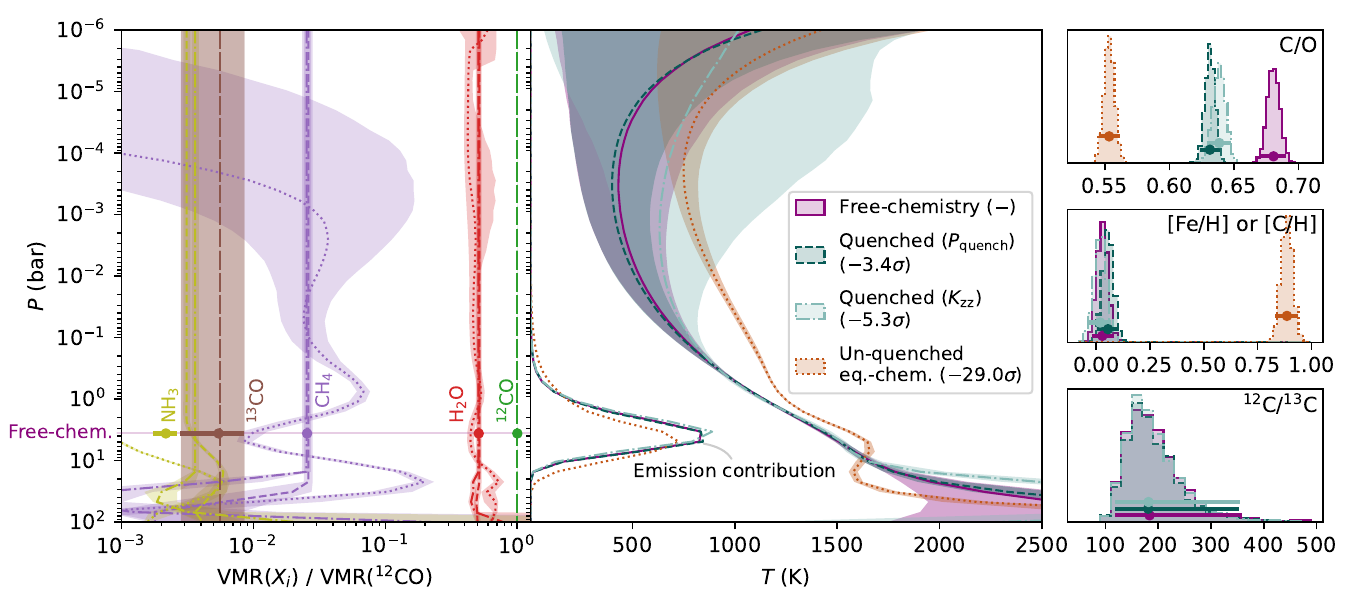}
    \caption{Retrieval results of three chemistry approaches: free-chemistry (solid/purple), quenched via $P_\mathrm{quench}$ (dashed/teal), quenched via $K_\mathrm{zz}$ (dot-dashed/turquoise) and un-quenched (dotted/orange) equilibrium chemistry. The linestyles are kept consistent in all panels and the envelopes and errorbars specify the $95\%$ confidence intervals. \textit{Left panel}: abundance profiles relative to $^{12}$CO for NH$_3$, $^{13}$CO, CH$_4$, H$_2$O and $^{12}$CO. The free-chemistry abundances are shown as markers with errorbars for clarity, but extend vertically over all pressures. \textit{Middle panel}: retrieved PT profiles and emission contribution functions. \textit{Right panels, from top to bottom}: posteriors of the C/O-ratio, metallicity, and carbon isotope ratio. The carbon abundance was used as a proxy for the metallicity (i.e. $[\mathrm{C/H}]$) in the free-chemistry results. The un-quenched equilibrium model finds substantially different results and is heavily disfavoured over the free-chemistry model at $29.0\sigma$.}
    \label{fig:chemistry_PT}
\end{figure*}

The constrained $^{13}$CO ($\log\ \mathrm{VMR}=-5.59^{+0.11}_{-0.13}$) and $^{12}$CO ($\log\ \mathrm{VMR}=-3.32^{+0.03}_{-0.03}$) abundances translate into an isotopologue ratio of $^{12}\mathrm{CO}/^{13}\mathrm{CO}=184^{+61}_{-40}$. The posterior of the carbon isotope ratio, along with the $95\%$ confidence interval, is shown in purple in the bottom right panel of Fig. \ref{fig:chemistry_PT}. Section \ref{sect:formation_discussion} provides an interpretation of the $\mathrm{^{12}C/^{13}C}$-ratio in regard to the possible formation history of DENIS J0255.

\subsection{Disequilibrium chemistry} \label{sect:diseq_chem}
As described in Sect. \ref{sect:chemistry}, we performed additional retrievals using the \texttt{pRT} chemical equilibrium implementation, both in its quenched and un-quenched form. The retrieval results of the four chemistry approaches are summarised in Table \ref{tab:params} and Fig. \ref{fig:chemistry_PT}. The left panel presents the abundance profiles relative to $^{12}$CO as high-resolution spectroscopy is most sensitive to abundance ratios. 

\begin{figure}[h!]
    \resizebox{\hsize}{!}{\includegraphics[width=17cm]{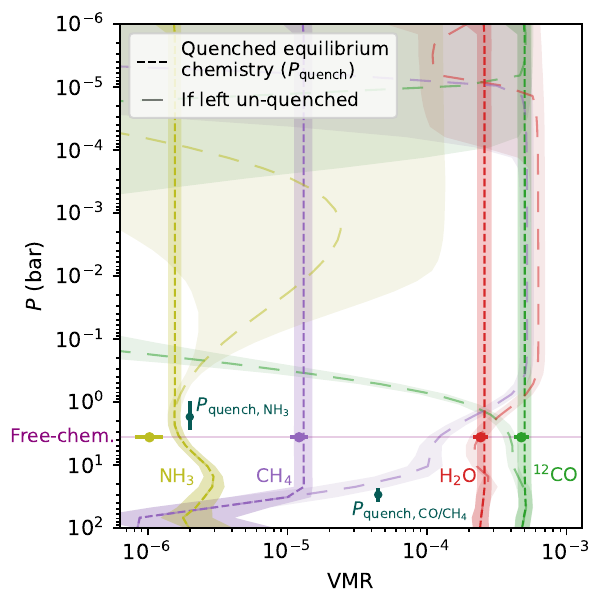}}
    \caption{Abundance profiles of $^{12}$CO, CH$_4$, H$_2$O, and NH$_3$, retrieved as part of the quenched (via $P_\mathrm{quench}$) chemical equilibrium retrieval. The short-dashed lines show the median quenched profiles used to compute the model spectra. For comparison, we show the un-quenched profiles with long dashes, indicating that CH$_4$ would become the dominant carbon-bearing species at pressures below $\sim$\,$1\ \mathrm{bar}$ in the case of chemical equilibrium. As in Fig. \ref{fig:chemistry_PT}, the shaded regions present the $95\%$ confidence envelopes.}
    \label{fig:VMR_profiles}
\end{figure}

The results from the $P_\mathrm{quench}$-retrieval show a good agreement with those obtained with the fiducial, free-chemistry model. The emission contribution and PT profile are consistent with the free-chemistry results, but the quenched-chemistry retrieval finds tighter temperature constraints below the photosphere due to its influence on the chemistry at higher altitudes. The CO-CH$_4$ network is quenched at a low altitude ($P_\mathrm{quench}(\mathrm{CO, CH_4, H_2O})=29.6^{+4.2}_{-3.6}\ \mathrm{bar}$) and thus the left panel of Fig. \ref{fig:chemistry_PT} presents almost identical abundances compared to the free-chemistry approach (markers with errorbars). The quench point of the N$_2$-NH$_3$ system is placed above that of CO-CH$_4$ ($P_\mathrm{quench}(\mathrm{NH_3})=1.7^{+0.5}_{-0.4}\ \mathrm{bar}$). We note that the photospheric abundance of NH$_3$ is somewhat higher ($\sim$\,$0.2\ \mathrm{dex}$) than found in the free-chemistry retrieval. The (median) absolute abundance profiles of $^{12}$CO, CH$_4$, H$_2$O, and NH$_3$ are shown as short-dashed lines in Fig. \ref{fig:VMR_profiles}. The long-dashed lines show the behaviour of the abundances if they were left un-quenched (i.e. in chemical equilibrium). At relevant pressures, the un-quenched NH$_3$ abundance profile is never as low as the retrieved free-chemistry abundance. The NH$_3$ quench point is therefore placed at the closest approximation. This discrepancy suggests an elemental nitrogen abundance lower than the solar value assumed for pRT's chemical equilibrium table. Given the non-detections of HCN and CO$_2$ in the fiducial model, it is not surprising that the respective quenching pressures remain unconstrained. A Bayesian evidence comparison reveals a slight preference for the free-chemistry model at $3.4\sigma$. 

Similar results are found with the retrieval implementing quenching via the $K_\mathrm{zz}$-coefficient. The retrieved abundances of CO, CH$_4$, and H$_2$O are in accordance with the free-chemistry and $P_\mathrm{quench}$-retrievals. However, the NH$_3$ mixing ratio is discrepant with the free-chemistry result ($\sim$\,$0.2\ \mathrm{dex}$). The photospheric temperature gradient is comparable to the previous models, but higher temperatures are generally retrieved above and below the peak contribution. This can likely be ascribed to the temperature-dependence of the vertical-mixing and chemical timescales. The retrieved $K_\mathrm{zz}=3.7^{+0.7}_{-0.6}\cdot10^9\ \mathrm{cm^2\ s^{-1}}$ can be converted to quench pressures for the different chemical systems. The derived quench points of the detected species ($P_{K_\mathrm{zz}}(\mathrm{CO, CH_4, H_2O})=17.4^{+0.9}_{-0.9}\ \mathrm{bar}$; $P_{K_\mathrm{zz}}(\mathrm{NH_3})=19.5^{+1.0}_{-1.0}\ \mathrm{bar}$) are inconsistent with those found in the $P_\mathrm{quench}$-retrieval. Evaluating the Bayes factor reveals a preference ($5.3\sigma$) for the free-chemistry model.

The un-quenched equilibrium retrieval is strongly disfavoured over the fiducial retrieval at $29.0\sigma$. The different choices for priors (e.g. log-uniform $\mathrm{VMR}$ vs. uniform $\mathrm{C/O}$-ratio) likely decreases the accuracy of the derived significance, but the conclusion that free-chemistry is heavily favoured over un-quenched chemical equilibrium should still hold. Figure \ref{fig:chemistry_PT} shows a PT profile with a minor inversion at the bottom of the photosphere and higher temperatures at altitudes above $10\ \mathrm{bar}$. Moreover, the un-quenched retrieval finds a super-solar metallicity ($\mathrm{[Fe/H]}=0.89^{+0.03}_{-0.03}$), a carbon-to-oxygen ratio ($\mathrm{C/O}=0.553^{+0.004}_{-0.004}$) that is reduced compared to the quenched- and free-chemistry results, and a surface gravity that hits the edge of the prior ($\log\textit{g}\sim$\,$6$). These peculiar results are the consequence of attempts to suppress the methane abundance and line depths with respect to $^{12}$CO. In the left panel of Fig. \ref{fig:chemistry_PT}, we find that the CH$_4$ abundance profile oscillates around the value retrieved with the quenched- and free-chemistry approaches. Maintaining a high temperature will favour CO as the primary carbon-bearing molecule and the retrieval adjusts the emission contribution function via the surface gravity and metallicity. In the absence of these adjustments, methane would become the dominant carbon-bearing molecule \citep{Zahnle_ea_2014,Moses_ea_2016} as demonstrated in Fig. \ref{fig:VMR_profiles}. The short-dashed lines indicate the median, quenched abundances profiles and the long-dashed lines show the abundances in the absence of quenching (i.e. in chemical equilibrium). We note that these un-quenched profiles are not found with the chemical equilibrium retrieval, as they would produce a poor fit to the observed spectrum. The low temperature between $\sim$\,$10^{-3}$ and $1\ \mathrm{bar}$ would convert CO into CH$_4$ (and H$_2$O) via the net reaction:
\begin{align}
    \mathrm{CO} + 3\ \mathrm{H}_2 \rightarrow \mathrm{CH}_4 + \mathrm{H}_2\mathrm{O}.
\end{align}
However, it is difficult to convert CO into CH$_4$ in reasonable timescales ($\tau_\mathrm{chem}\sim$\,$10^8\ \mathrm{s}$ at $1\ \mathrm{bar}$ and $1250\ \mathrm{K}$; \citealt{Cooper_ea_2006}) due to its strong triple bond. As a consequence, vertical mixing can retain a CO-rich atmosphere even at low temperatures. The inferred CO and CH$_4$ abundances are therefore evidence of a chemical disequilibrium in the atmosphere of DENIS J0255. The relation between the retrieved disequilibrium and the atmospheric dynamics is discussed in Sect. \ref{sect:quenching_discussion}.

\begin{figure*}[h!]
    \centering
    \includegraphics[width=17cm]{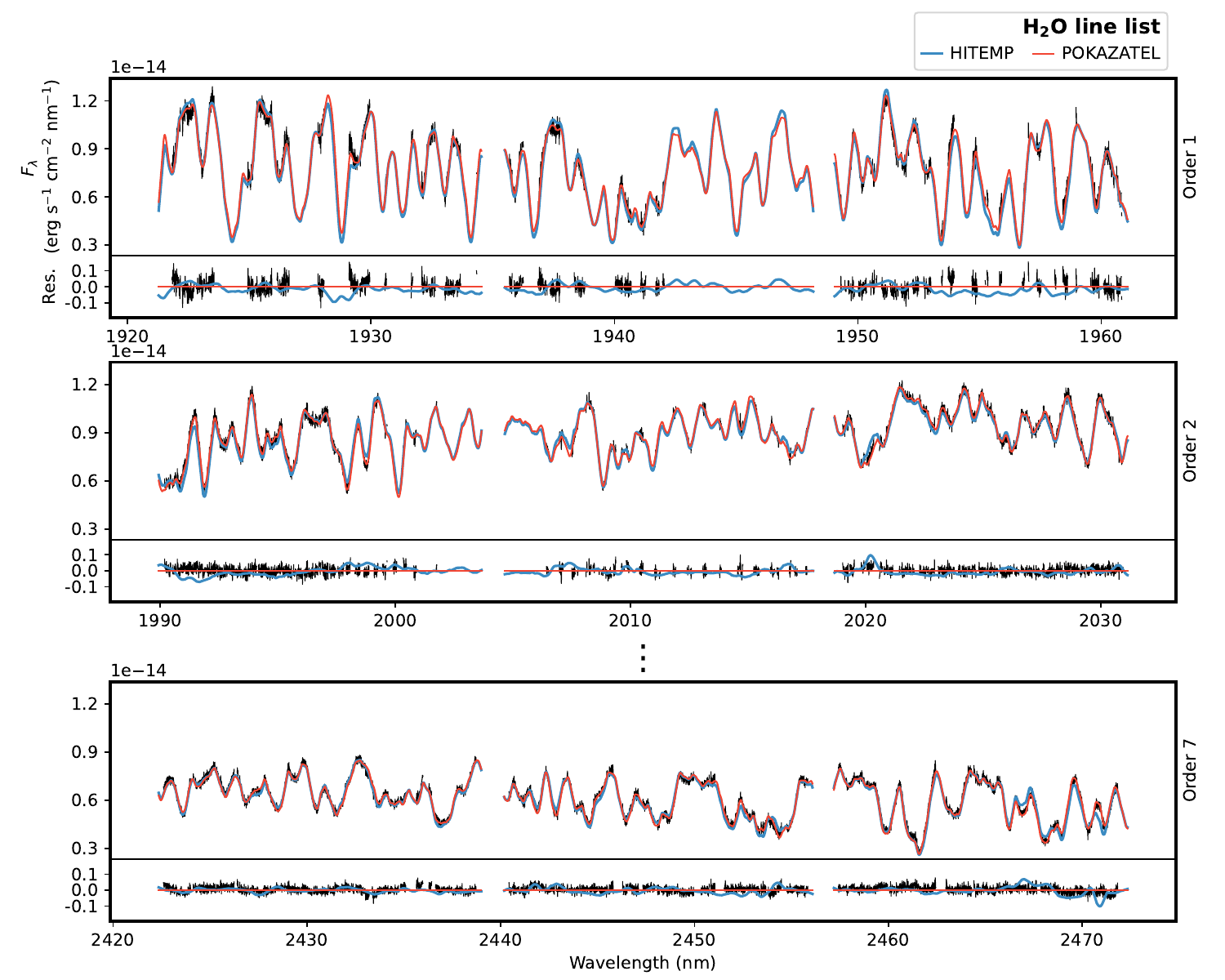}
    \caption{Comparison between the best-fitting model spectra using the HITEMP (blue; \citealt{Rothman_ea_2010}) and ExoMol (orange; \citealt{Polyansky_ea_2018}) line lists for H$_2$O. Only the 1$^\mathrm{st}$, 2$^\mathrm{nd}$ and 7$^\mathrm{th}$ spectral orders of the K2166 wavelength setting are shown since the other orders showed minor discrepancies. The residuals of the observed spectrum are shown with respect to the fiducial model which utilises the ExoMol line list. }
    \label{fig:H2O_line_list}
\end{figure*}

As shown in the top right panel of Fig. \ref{fig:chemistry_PT}, the free-chemistry model obtains a carbon-to-oxygen ratio of $\mathrm{C/O}=0.681^{+0.005}_{-0.005}$. This atmospheric $\mathrm{C/O}$-ratio might be inflated due to the condensation of oxygen into silicates \citep{Burrows_ea_1999,Line_ea_2015,Calamari_ea_2022}, which could explain its super-solar value ($\mathrm{C/O}_\odot=0.59\pm0.08$; \citealt{Asplund_ea_2021}). The quenched equilibrium models include this oxygen sequestration and thus retrieve lower carbon-to-oxygen ratios ($\mathrm{C/O}_\mathrm{quench}=0.631^{+0.004}_{-0.004}$; $\mathrm{C/O}_{K_\mathrm{zz}}=0.639^{+0.005}_{-0.005}$), despite the similar gaseous ratios visible from the relative abundance profiles in the left panel of Fig. \ref{fig:chemistry_PT}. While it is more difficult to constrain absolute abundances due to the degeneracies seen in Fig. \ref{fig:fiducial_corner}, high-resolution spectroscopy is very sensitive to abundance ratios, thus resulting in smaller uncertainties for the $\mathrm{C/O}$-ratio. However, we reiterate that the uncertainties are likely under-estimated due to the chosen constant sampling efficiency of $5\%$ \citep{Chubb_ea_2022}. For the free-chemistry model, we employ the carbon abundance relative to hydrogen as a proxy for the atmospheric metallicity. The metallicity retrieved in the free-chemistry ($[\mathrm{C/H}]=0.03^{+0.03}_{-0.03}$) and quenched equilibrium retrievals ($[\mathrm{Fe/H}]_\mathrm{quench}=0.06^{+0.02}_{-0.03}$; $[\mathrm{Fe/H}]_{K_\mathrm{zz}}=0.02^{+0.03}_{-0.03}$) are broadly consistent with the solar measurement, but their difference is observed as a small shift in the absolute abundances of Fig. \ref{fig:VMR_profiles}. As depicted in the lower right panel of Fig. \ref{fig:chemistry_PT}, the CO isotopologue ratios of the quenched equilibrium models ($\mathrm{^{12}CO/^{13}CO}_\mathrm{quench}=181^{+57}_{-37}$; $\mathrm{^{12}CO/^{13}CO}_{K_\mathrm{zz}}=182^{+61}_{-37}$) are in agreement with that of the free-chemistry retrieval ($\mathrm{^{12}CO/^{13}CO}=186^{+61}_{-40}$). Evidently, the un-quenched equilibrium model fails to constrain the carbon isotope ratio due to its poor fit to the observed spectrum. 

\subsection{H$_\textit{2}$O line list validation} \label{sect:H2O_line_list}
High-resolution spectra, like the one analysed in this work, enable reliability tests of the utilised molecular line lists. Accurate and complete line lists improve the robustness of detections of chemical species in exoplanet and brown dwarf atmospheres, as well as improving the modelled abundances, temperature profiles, etc. In this section, we make a comparison between the HITEMP \citep{Rothman_ea_2010} and ExoMol \citep{Polyansky_ea_2018} line lists for H$_2$O. Since ExoMol's POKAZATEL line list was employed for the fiducial model, a separate retrieval was carried out using the HITEMP line list. Figure \ref{fig:H2O_line_list} presents the difference in best-fitting models for the two bluest and the reddest spectral orders of the K2166 wavelength setting. The four remaining orders showed negligible discrepancies and are therefore not displayed. The HITEMP line list exhibits considerable deviations from the observed spectrum which results in a poor fit in certain regions of orders 1, 2 and 7. These orders are also the most H$_2$O-dominated at the blue and red edges of the K-band. Furthermore, the Bayesian evidence comparison shows a strong preference for the ExoMol line list at $25.2\sigma$.

As indicated in Table \ref{tab:params}, the HITEMP retrieval results are discrepant from those obtained with the fiducial model which uses the ExoMol H$_2$O line list. For instance, the increased GP amplitudes for orders 1, 2 and 7 demonstrate that these wavelength regions are poorly fit as a consequence of inaccurate or incomplete H$2$O opacities in the HITEMP line list. Moreover, the surface gravity and metallicity are higher in the HITEMP retrieval and it has a shallower temperature gradient. We note that the retrieved isotopologue ratio ($\mathrm{^{12}CO/^{13}CO}=118^{+27}_{-20}$) is reduced because the retrieval compensates for the inadequate fit with stronger $^{13}$CO absorption in the reddest order. Hence, this comparison highlights the importance of accurate line lists for the study of weak features such as isotopologues. 

\section{Discussion}
\subsection{Validity of cloud-free solution}\label{sect:cloud_effects}
We find cloud-free solutions for each retrieval, regardless of the employed chemical model. A possible explanation could be that the fast rotation obscures the continuum by broadening the spectral line wings. This makes it more difficult to probe the cloud layers that are already expected to reside at low altitudes due to the cool temperature of this late L-type object. In the upper right panel of Fig. \ref{fig:fiducial_corner}, we show the condensation curves of several cloud species obtained from \citet{Visscher_ea_2006} (MnS) and \citet{Visscher_ea_2010} (MgSiO$_3$, Mg$_2$SiO$_4$, Fe). We find that Mg$_2$SiO$_4$ and Fe are presumably condensing out below the K-band photosphere. However, the intersection of the PT profile with the condensation curves of MnS and MgSiO$_3$ suggests that these species could form clouds at pressures probed by the observed CRIRES$^+$ spectrum, but this expected cloud opacity is either negligible or is not sufficiently traced by our grey cloud model. Other retrieval studies of high-resolution K-band spectra have similarly resulted in cloud-free solutions (e.g. \citealt{Zhang_ea_2021b,Xuan_ea_2022,Landman_ea_2024}). The K-band is expected to be less sensitive to clouds than the shorter wavelengths of the J- and H-band. The opacity of gaseous molecules generally increases towards longer wavelengths, leading to higher altitudes becoming probed in the K-band \citep{Marley_ea_2002}. Conversely, the opacity of small, sub-micron cloud particles decreases with wavelength \citep{Morley_ea_2014,Marley_ea_2015}, thereby hampering attempts to constrain the cloud opacity with K-band spectra.

\begin{figure}[h!]
    \resizebox{\hsize}{!}{\includegraphics[width=17cm]{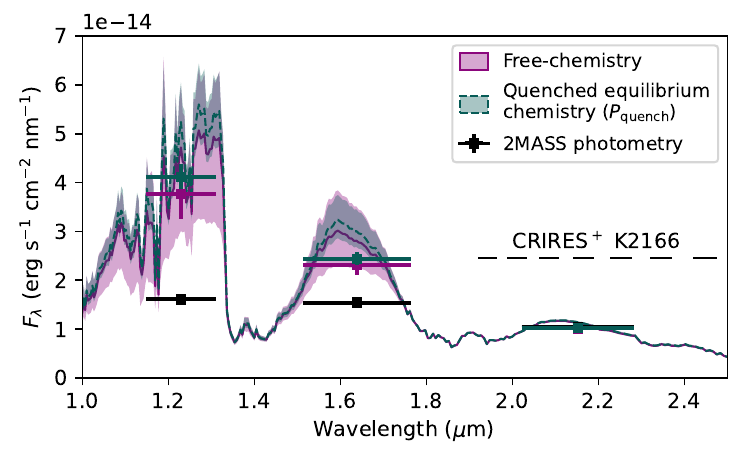}}
    \caption{Comparison between the retrieved, cloud-free model spectra and the 2MASS J-, H-, and Ks-band photometry \citep{Cutri_ea_2003}. The shaded regions show the $95\%$ confidence intervals.}
    \label{fig:low_res_comparison}
\end{figure}

Figure \ref{fig:low_res_comparison} compares the 2MASS J-, H-, and Ks-band photometry ($\mathrm{J}=13.25\pm0.03$, $\mathrm{H}=12.20\pm0.02$, $\mathrm{Ks}=11.56\pm0.02\ \mathrm{mag}$; \citealt{Cutri_ea_2003}) with the retrieved model spectra, where equilibrium abundances were adopted for FeH, H$_2$S, VO, TiO, K, and Na based on the retrieved $\mathrm{C/O}$-ratio and metallicity. We note that the quenched equilibrium retrieval in Fig. \ref{fig:low_res_comparison} shows a narrower flux envelope at shorter wavelengths due to its tighter temperature constraint at low altitudes (see Fig. \ref{fig:chemistry_PT}) compared to the free-chemistry results. The model spectra are consistent with the Ks-band photometry on account of the simple flux calibration (Sect. \ref{sect:obs_and_reduction}), but show excess flux in J- and H-band. The true J- and H-band fluxes are likely reduced by clouds that are present below the K-band photosphere and have a decreased opacity at the wavelengths of the CRIRES$^+$ spectrum. As a consequence of the high flux at short wavelengths, we obtain effective temperature estimates of $T_\mathrm{eff}=1554^{+263}_{-53}\ \mathrm{K}$ and $1723^{+262}_{-141}\ \mathrm{K}$ for the free-chemistry and quenched equilibrium ($P_\mathrm{quench}$) retrievals, respectively. Evidently, the absence of short-wavelength cloud opacity raises the effective temperature to be inconsistent with the $T_\mathrm{eff}\sim$\,$1400\ \mathrm{K}$ found using cloudy models (e.g. \citealt{Cushing_ea_2008,Charnay_ea_2018,Lueber_ea_2022}). Therefore, the lack of constraints on a grey cloud in this work should not be seen as evidence of a cloud-free atmosphere for DENIS J0255 as that would be incompatible with measurements at other wavelengths.

\subsection{Chemical quenching \& vertical mixing}\label{sect:quenching_discussion}
We find conflicting quench points between the quenched equilibrium retrievals. The $P_\mathrm{quench}$-retrieval quenches the N$_2$-NH$_3$ system above the CO-CH$_4$ quench point. However, for these conditions, the chemical timescale of NH$_3$ is expected to be longer than CO-CH$_4$ and thus vertical mixing should already dominate the N$_2$-NH$_3$ system at lower altitudes \citep{Zahnle_ea_2014}. As discussed in Sect. \ref{sect:diseq_chem}, NH$_3$ is quenched at the lowest abundance permitted by the PT profile, leading to the anomalous quench pressure. We find the expected order of quench points in the $K_\mathrm{zz}$-retrieval because the chemical timescale prescription of \citet{Zahnle_ea_2014} was adopted.

Following the method described by \citet{Miles_ea_2020} and \citet{Xuan_ea_2022}, we can convert the constrained quench points from the $P_\mathrm{quench}$-retrieval into vertical eddy diffusion coefficients, $K_\mathrm{zz}$. Since the N$_2$-NH$_3$ network is likely quenched at erroneously high altitudes, we only consider the CO-CH$_4$ system. At the quench point of this network, the chemical reaction timescale ($\tau_\mathrm{CO-CH_4}=5.8^{+3.4}_{-2.3}\ \mathrm{s}$; Eq. 14 of \citealt{Zahnle_ea_2014}) is set equal to the mixing timescale. Using the inverse of Eq. \ref{eq:tau_mix}, we find an eddy diffusion coefficient of $K_\mathrm{zz,\ CO-CH_4}=2.7^{+2.1}_{-1.0}\cdot10^{10}\ \mathrm{cm^2\ s^{-1}}$ (with $L=H$). For comparison, the directly-retrieved diffusion coefficient is constrained about an order of magnitude lower, at $K_\mathrm{zz}=3.7^{+0.7}_{-0.6}\cdot10^9\ \mathrm{cm^2\ s^{-1}}$. However, it is difficult to compare the coefficients as the $K_\mathrm{zz}$-retrieval is also affected by the NH$_3$ abundance that is required to provide a good fit.

Assuming full convection, the upper limit from mixing length theory \citep{Gierasch_ea_1985,Zahnle_ea_2014} is $K_\mathrm{zz}\lesssim1.1\cdot10^9\ \mathrm{cm^2\ s^{-1}}$ for DENIS J0255, using $T_\mathrm{eff}=1400\ \mathrm{K}$ and $\log\textit{g}=5.3$. Therefore, the inferred $K_\mathrm{zz,\ CO-CH_4}$ exceeds the upper bound, assuming that the mixing length is equal to the scale height ($L=H$). For a scaling factor of $\alpha=0.2$ (i.e. $L=0.2H$), the derived diffusion coefficient $K_\mathrm{zz,\ CO-CH_4}$ equals the convective upper limit. Analogous to the analysis presented by \citet{Xuan_ea_2022}, this suggests that the vertical mixing of DENIS J0255 is primarily driven by efficient convection. The diffusivity derived for DENIS J0255 is similar to that of \citet{Xuan_ea_2022} who find $K_\mathrm{zz}=5\cdot10^{10}-1\cdot10^{14}\ \mathrm{cm^2\ s^{-1}}$ ($L=H$) or $5\cdot10^8-1\cdot10^{12}\ \mathrm{cm^2\ s^{-1}}$ ($L=0.1H$) for the brown dwarf companion HD 4747B. It is encouraging to find agreement in the inferred $K_\mathrm{zz}$ between HD 4747B and DENIS J0255 as the two brown dwarfs share similar effective temperatures, surface gravities and ages.

In addition to upsetting chemical equilibrium, vertical mixing affects the cloud opacity by transporting condensable vapour upwards to cooler temperatures where condensation can occur. Moreover, vertical mixing counteracts the gravitational settling of cloud particles, thus allowing clouds to be present at higher altitudes (e.g. \citealt{Ormel_ea_2019,Mukherjee_ea_2022}). Hence, the high diffusivity, $K_\mathrm{zz}$, found for DENIS J0255 suggests an increased cloud thickness \citep{Ormel_ea_2019}. If our cloud-free solution is truly evidence for the absence of any detectable cloud opacity in the K-band, the obtained eddy diffusion coefficient might help in constraining cloud properties such as the particle sizes. Furthermore, we note that $K_\mathrm{zz}$ could decrease with altitude (e.g. \citealt{Charnay_ea_2018,Tan_ea_2019,Tan_2022}) which might also prevent the clouds from reaching into the probed layers.

\subsection{Formation history \& isotope ratios}\label{sect:formation_discussion}
The fiducial, free-chemistry retrieval constrains the CO isotopologue ratio at $^{12}\mathrm{CO}/^{13}\mathrm{CO}=184^{+61}_{-40}$. Despite the tentative nature of the $^{13}$CO signal, isotope ratios lower than $\mathrm{^{12}C/^{13}C}\lesssim$\,$97$ ($0.15$-th percentile) appear incompatible with the weak $^{13}$CO absorption in the observed spectrum. Hence, DENIS J0255 is likely depleted in $^{13}$C compared to the local ISM ($\mathrm{^{12}C/^{13}C}=68\pm15$; \citealt{Milam_ea_2005}), akin to earlier work for the young brown dwarf 2M 0355 ($108\pm10$; \citealt{Zhang_ea_2021b,Zhang_ea_2022}) and the fully convective M dwarf binary GJ 745 ($296\pm45$, $224\pm26$; \citealt{Crossfield_ea_2019}). A potential explanation for the measured isotopologue ratio is that DENIS J0255 was formed from a parent cloud under-abundant in $^{13}$CO, similar to several molecular clouds and young stellar objects in the solar neighbourhood ($\mathrm{^{12}C/^{13}C}\sim$\,$167$; \citealt{Lambert_ea_1994}, $\sim$\,$125$; \citealt{Federman_ea_2003}, $\sim$\,$86-158$; \citealt{Goto_ea_2003}, $\sim$\,$85-165$; \citealt{Smith_ea_2015}). Due to its expected mature age ($2$-$4\ \mathrm{Gyr}$; \citealt{Creech-Eakman_ea_2004}), the parent cloud of DENIS J0255 possibly did not experience considerable $^{13}$C-enrichment by asymptotic giant branch stars, in contrast to the present-day ISM \citep{Milam_ea_2005,Romano_ea_2017}. With the caveat that $^{13}$CO is only weakly identified, this work further supports a possible distinction between the formation pathways of isolated brown dwarfs and super-Jupiters as proposed by \citet{Zhang_ea_2021b}. 

The quenched equilibrium models constrain the isotopologue ratio to $\mathrm{^{12}CO/^{13}CO}_\mathrm{quench}=181^{+57}_{-37}$ and $\mathrm{^{12}CO/^{13}CO}_{K_\mathrm{zz}}=182^{+61}_{-37}$, in line with the ratio found by the free-chemistry model. In contrast, another proposed tracer of planet formation and evolution, the $\mathrm{C/O}$-ratio, results in discrepancies between the models, likely due to the condensation of oxygen into silicate clouds \citep{Burrows_ea_1999,Line_ea_2015,Calamari_ea_2022}. The similarity in the retrieved $\mathrm{^{12}C/^{13}C}$-ratios could be an indication that isotope ratios are less sensitive to cloud-condensation and dis-equilibrium chemistry compared to the $\mathrm{C/O}$-ratio. Such a reduced model-dependency forms a good argument for pursuing isotope ratios as tracers of planet formation and evolution.

\section{Conclusions}\label{sect:conclusions}
We analysed the high-resolution K-band spectrum of the late L9-dwarf DENIS J0255, observed with CRIRES$^+$ as part of the ESO SupJup Survey. Employing different chemical models in combination with atmospheric retrievals, we find that a free-chemistry approach produces near-identical results to a model where the chemical equilibrium abundances are quenched above certain altitudes. The retrieval results of an un-quenched equilibrium model are heavily disfavoured at $29.0\sigma$ (based on a Bayesian model comparison), thereby unveiling a detection of chemical dis-equilibrium. In equilibrium, CH$_4$ would be the primary carbon-bearing molecule at the temperature of this brown dwarf, but the long chemical timescales enable DENIS J0255 to retain a CO-rich atmosphere via vertical mixing. 

The H$_2$O-dominated edges of the K-band are poorly fit with the HITEMP H$_2$O line list, in contrast to ExoMol's POKAZATEL line list. The Bayesian evidences yield a $25.2\sigma$ preference for the ExoMol line list. The comparison reveals discrepancies between the retrieved parameters, thus emphasising the importance of accurate and complete line lists for reliable results. 

A Bayesian evidence comparison shows that the inclusion of CH$_4$, NH$_3$, and the $^{13}$CO isotopologue is favoured at a significance of $23.1\sigma$, $7.9\sigma$, and $2.7\sigma$, respectively. A cross-correlation analysis results in detection significances of $30.9\sigma$ (CH$_4$), $10.6\sigma$ (NH$_3$), and $3.7\sigma$ ($^{13}$CO). Apart from CH$_4$, we report the first confident detection of NH$_3$ in the spectrum of DENIS J0255. The retrieved abundance of the weakly-detected $^{13}$CO results in an isotopologue ratio of $^{12}\mathrm{CO}/^{13}\mathrm{CO}=184^{+61}_{-40}$ while a lower limit of $\gtrsim$\,$97$ can be established. The high ratio suggests that DENIS J0255 is depleted in $^{13}$C compared to the present-day, local ISM. Since DENIS J0255 is expected to be several gigayears old, it is expected that its parent cloud was less enriched in $^{13}$C from evolved stars. As the second measurement of the CO isotopologue ratio in a brown dwarf's atmosphere, this paper lends further credence to a possible distinction between the formation pathways of isolated brown dwarfs and super-Jupiters. Future work as part of the ESO SupJup Survey will assess whether depletions of $^{13}$C are more commonly seen for these isolated objects and whether the carbon isotope ratio can truly distinguish their formation scenario from that of planetary-mass companions. 

\begin{acknowledgements}
S.d.R. and I.S. acknowledge funding from the European Research Council (ERC) under the European Union's Horizon 2020 research and innovation program under grant agreement No 694513. 
A.S.L. acknowledges financial support from the Severo Ochoa grant CEX2021-001131-S funded by MCIN/AEI/ 10.13039/501100011033. 
T.S. acknowledges the support from the Netherlands Organisation for Scientific Research (NWO) through grant VI.Veni.202.230.
This work used the Dutch national e-infrastructure with the support of the SURF Cooperative using grant no. EINF-4556. 
\end{acknowledgements}

\bibliographystyle{aa}
\bibliography{references.bib}

\begin{appendix}\onecolumn
\begin{landscape}
    \section{SupJup observations}
    {\setlength{\tabcolsep}{3.5pt}
{\scriptsize
\begin{longtable}{ll|llllllll|rrrrr}
\caption{Targets observed as part of the ESO SupJup Survey (Program ID: 1110.C-4264, PI: Snellen), sorted by right ascension. \label{tab:observed_targets}} \\

\hline\hline
\textbf{System Name} & \textbf{On-Slit Targets} & $\mathbf{d}$ (pc) & \textbf{BANYAN $\mathbf{\Sigma}$}$^{(1)}$ & \textbf{Age} (Myr) & \textbf{SpType} & $\mathrm{\mathbf{m_J}}$ (mag) & $\mathrm{\mathbf{m_K}}$ (mag) & $\mathbf{M}$ ($M_\mathrm{Jup}$) & $\mathbf{a}$ ("/AU) & \textbf{Obs. Date} & \textbf{Setting} & \textbf{AO} & \textbf{Slit} & \textbf{Exp. Time} \\
\hline
\hline
\endfirsthead

\caption{Continued.} \\

\hline\hline
\textbf{System Name} & \textbf{On-Slit Targets} & $\mathbf{d}$ (pc) & \textbf{BANYAN $\mathbf{\Sigma}$}$^{(1)}$ & \textbf{Age} (Myr) & \textbf{SpType} & $\mathrm{\mathbf{m_J}}$ (mag) & $\mathrm{\mathbf{m_K}}$ (mag) & $\mathbf{M}$ ($M_\mathrm{Jup}$) & $\mathbf{a}$ ("/AU) & \textbf{Obs. Date} & \textbf{Setting} & \textbf{AO} & \textbf{Slit} & \textbf{Exp. Time} \\
\hline
\hline
\endhead

\hline
\endfoot

HD 1160 & \begin{tabular}{@{}l}HD 1160B, \\HD 1160C\end{tabular} & 120.9$\pm$0.5$^{(2)}$ & FIELD (99.9\%) & 10-20$^{(3)}$ & \begin{tabular}{@{}l}M5.5$^{(4)}$, \\M3.5$^{(5)}$\end{tabular} & \begin{tabular}{@{}l}15.83$\pm$0.10, \\13.31$\pm$0.04$^{(5)}$\end{tabular} & \begin{tabular}{@{}l}14.12$\pm$0.05, \\12.18$\pm$0.06$^{(5)}$\end{tabular} & \begin{tabular}{@{}l}20$^{(3)}$, \\231$\pm$42$^{(5)}$\end{tabular} & \begin{tabular}{@{}l}0.78/80, \\5.15/530$^{(5)}$\end{tabular} & 2022 11 01 & K2166 & Yes & 0.4 & 45$\times$300s = 3h45m \\ \hline
LSPM J0036+1821 & LSPM J0036 & 8.74$\pm$0.01$^{(2)}$ & FIELD (99.9\%) & - & L3.5$^{(6)}$ & 12.47$\pm$0.03$^{(7)}$ & 11.06$\pm$0.02$^{(7)}$ & - & - & 2022 11 03 & K2166 & No & 0.2 & 10$\times$300s = 0h50m \\
 &  &  &  &  &  &  &  &  &  & 2022 11 04 & J1226 & No & 0.2 & 14$\times$300s = 1h10m \\ \cline{15-15} 
 &  &  &  &  &  &  &  &  &  &  &  &  &  & 2h00m \\ \hline
2MASS J01033563-5515561 & \begin{tabular}{@{}l}2M 0103AB, \\2M 0103(AB)b\end{tabular} & 12.86$\pm$1.97$^{(8)}$ & THA (99.3\%) & 45$\pm$4$^{(9)}$ & \begin{tabular}{@{}l}M5.5, \\L$^{(10)}$\end{tabular} & \begin{tabular}{@{}l}10.16$\pm$0.02$^{(7)}$, \\15.4$\pm$0.3$^{(10)}$\end{tabular} & \begin{tabular}{@{}l}9.24$\pm$0.02$^{(7)}$, \\13.6$\pm$0.2$^{(10)}$\end{tabular} & \begin{tabular}{@{}l}199/178$\pm$21, \\13$\pm$1$^{(10)}$\end{tabular} & 1.72/84$^{(10)}$ & 2022 11 04 & K2166 & No & 0.4 & 34$\times$300s = 2h50m \\ \hline
2MASS J01225093-2439505 & \begin{tabular}{@{}l}2M 0122A, \\2M 0122B\end{tabular} & 33.74$\pm$0.03$^{(2)}$ & ABDMG (98.4\%) & 149$^{+51}_{-19}$$^{(9)}$ & \begin{tabular}{@{}l}M3.5$^{(11)}$, \\L4-6$^{(12)}$\end{tabular} & \begin{tabular}{@{}l}10.02$\pm$0.03, \\16.81$\pm$0.14$^{(12)}$\end{tabular} & \begin{tabular}{@{}l}9.17$\pm$0.03, \\14.53$\pm$0.05$^{(12)}$\end{tabular} & \begin{tabular}{@{}l}419$\pm$52, \\23-27$^{(12)}$\end{tabular} & 1.45/52$\pm$6$^{(12)}$ & 2022 11 02 & K2166 & Yes & 0.4 & 49$\times$300s = 4h05m \\
 &  &  &  &  &  &  &  &  &  & 2022 11 03 & K2166 & Yes & 0.4 & 42$\times$300s = 3h30m \\ \cline{15-15} 
 &  &  &  &  &  &  &  &  &  &  &  &  &  & 7h35m \\ \hline
2MASS J01415823-4633574 & 2M 0141 & 36.65$\pm$0.59$^{(2)}$ & THA (99.9\%) & 45$\pm$4$^{(9)}$ & L0$^{(13)}$ & 14.83$\pm$0.04$^{(7)}$ & 13.10$\pm$0.03$^{(7)}$ & - & - & 2023 01 03 & K2166 & No & 0.4 & 12$\times$300s = 1h00m \\ \hline
2MASS J02192210-3925225 & \begin{tabular}{@{}l}2M 0219A, \\2M 0219B\end{tabular} & 40.2$\pm$0.1$^{(2)}$ & THA (99.9\%) & 45$\pm$4$^{(9)}$ & \begin{tabular}{@{}l}M6, \\L4$^{(14)}$\end{tabular} & \begin{tabular}{@{}l}11.32$\pm$0.03, \\15.54$\pm$0.10$^{(14)}$\end{tabular} & \begin{tabular}{@{}l}10.44$\pm$0.03, \\13.82$\pm$0.10$^{(14)}$\end{tabular} & \begin{tabular}{@{}l}113$\pm$12, \\13.9$\pm$1.1$^{(14)}$\end{tabular} & 3.96/156$\pm$10$^{(14)}$ & 2022 11 01 & K2166 & No & 0.4 & 26$\times$300s = 2h10m \\
 &  &  &  &  &  &  &  &  &  & 2022 11 03 & K2166 & No & 0.4 & 16$\times$300s = 1h20m \\
 &  &  &  &  &  &  &  &  &  & 2022 11 04 & K2166 & No & 0.4 & 32$\times$300s = 2h40m \\ \cline{15-15} 
 &  &  &  &  &  &  &  &  &  &  &  &  &  & 6h10m \\ \hline
DENIS J025503.3-470049 & DENIS J0255 & 4.868$\pm$0.004$^{(2)}$ & FIELD (99.9\%) & - & L9$^{(15)}$ & 13.24$\pm$0.03$^{(7)}$ & 11.56$\pm$0.02$^{(7)}$ & - & - & 2022 11 02 & K2166 & No & 0.2 & 12$\times$300s = 1h00m \\ \hline
2MASS J03552337+1133437 & 2M 0355 & 9.16$\pm$0.04$^{(2)}$ & ABDMG (99.9\%) & 149$^{+51}_{-19}$$^{(9)}$ & L5$^{(13)}$ & 14.05$\pm$0.02$^{(7)}$ & 11.53$\pm$0.02$^{(7)}$ & - & - & 2022 11 02 & K2166 & No & 0.2 & 10$\times$300s = 0h50m \\ \hline
FU Tau & \begin{tabular}{@{}l}FU Tau A, \\FU Tau B\end{tabular} & 127.6$\pm$1.2$^{(2)}$ & TAU (96.6\%) & 1-2$^{(16)}$ & \begin{tabular}{@{}l}M7.25, \\M9.25$^{(17)}$\end{tabular} & \begin{tabular}{@{}l}10.78$\pm$0.03, \\15.04$\pm$0.12$^{(7)}$\end{tabular} & \begin{tabular}{@{}l}9.32$\pm$0.02, \\13.33$\pm$0.10$^{(7)}$\end{tabular} & \begin{tabular}{@{}l}52.4, \\15.7$^{(17)}$\end{tabular} & 5.7/800$^{(17)}$ & 2023 01 01 & K2166 & No & 0.4 & 24$\times$300s = 2h00m \\ \hline
DH Tau & \begin{tabular}{@{}l}DH Tau A, \\DH Tau B\end{tabular} & 133.4$\pm$0.5$^{(2)}$ & TAU (99.8\%) & 1-2$^{(16)}$ & \begin{tabular}{@{}l}M0.5, \\L2$^{(18)}$\end{tabular} & \begin{tabular}{@{}l}9.77$\pm$0.02, \\15.71$\pm$0.05$^{(18)}$\end{tabular} & \begin{tabular}{@{}l}8.18$\pm$0.03, \\14.19$\pm$0.02$^{(18)}$\end{tabular} & \begin{tabular}{@{}l}670$\pm$42$^{(19)}$, \\8-22$^{(20)}$\end{tabular} & 2.34/328.2$\pm$0.4$^{(21)}$ & 2022 12 31 & K2166 & No & 0.4 & 16$\times$300s = 1h20m \\
 &  &  &  &  &  &  &  &  &  & 2023 01 01 & K2166 & No & 0.4 & 14$\times$300s = 1h10m \\ \cline{15-15} 
 &  &  &  &  &  &  &  &  &  &  &  &  &  & 2h30m \\ \hline
2MASS J04341527+2250309 & 2M 0434 & 163$\pm$10$^{(2)}$ & TAU (87.8\%) & 1-2$^{(16)}$ & M7$^{(22)}$ & 13.74$\pm$0.03$^{(7)}$ & 11.85$\pm$0.02$^{(7)}$ & - & - & 2022 12 31 & K2166 & No & 0.4 & 1$\times$300s = 0h05m \\
 &  &  &  &  &  &  &  &  &  & 2023 02 01 & K2166 & No & 0.4 & 22$\times$300s = 1h50m \\
 &  &  &  &  &  &  &  &  &  & 2023 02 01 & J1226 & No & 0.4 & 10$\times$300s = 50m \\ \cline{15-15} 
 &  &  &  &  &  &  &  &  &  &  &  &  &  & 2h45m \\ \hline
2MASS J05002100+0330501 & 2M 0500 & 13.23$\pm$0.05$^{(2)}$ & FIELD (99.9\%) & - & L4$^{(23)}$ & 13.67$\pm$0.02$^{(7)}$ & 12.06$\pm$0.02$^{(7)}$ & - & - & 2023 02 27 & K2166 & No & 0.4 & 18$\times$300 = 1h30m \\ \hline
2MASS J05012406-0010452 & 2M 0501 & 20.9$\pm$0.3$^{(2)}$ & FIELD (99.9\%) & - & L4$^{(13)}$ & 14.98$\pm$0.04$^{(7)}$ & 12.96$\pm$0.04$^{(7)}$ & - & - & 2022 12 31 & K2166 & No & 0.4 & 3$\times$300s = 0h15m \\
 &  &  &  &  &  &  &  &  &  & 2023 02 01 & K2166 & No & 0.4 & 12$\times$300s = 1h00m \\ \cline{15-15} 
 &  &  &  &  &  &  &  &  &  &  &  &  &  & 1h15m \\ \hline
2MASS J05233822-1403022 & 2M 0523 & 12.73$\pm$0.02$^{(2)}$ & FIELD (99.9\%) & - & L2.5$^{(24)}$ & 13.08$\pm$0.02$^{(7)}$ & 11.638$\pm$0.03$^{(7)}$ & - & - & 2023 01 03 & K2166 & No & 0.4 & 12$\times$300s = 1h00m \\ \hline
$\beta$ Pic & $\beta$ Pic b & 19.44$\pm$0.05$^{(2)}$ & $\beta$PMG (99.9\%) & 24$\pm$3$^{(9)}$ & L2$^{(25)}$ & 14.00$\pm$0.08$^{(25)}$ & 12.30$\pm$0.15$^{(25)}$ & 11.90$^{+2.93}_{-3.04}$$^{(26)}$ & $\sim$0.5/10.6$\pm$0.5$^{(27)}$ & 2023 01 02 & K2166 & Yes & 0.2 & 210$\times$120s = 7h00m \\ \hline
2MASS J05591914-1404488 & 2M 0559 & 10.50$\pm$0.08$^{(2)}$ & FIELD (99.9\%) & - & T4.5$^{(28)}$ & 13.80$\pm$0.02$^{(7)}$ & 13.58$\pm$0.05$^{(7)}$ & - & - & 2023 02 01 & K2166 & No & 0.4 & 16$\times$300s = 1h20m \\
 &  &  &  &  &  &  &  &  &  & 2023 02 01 & J1226 & No & 0.4 & 4$\times$300s = 0h20m \\ \cline{15-15} 
 &  &  &  &  &  &  &  &  &  &  &  &  &  & 1h40m \\ \hline
DENIS J060852.8-275358 & 2M 0608 & 44.2$\pm$0.3$^{(2)}$ & FIELD (99.9\%) & - & M9.6$^{(29)}$ & 13.60$\pm$0.03$^{(7)}$ & 12.37$\pm$0.03$^{(7)}$ & - & - & 2023 02 26 & K2166 & No & 0.4 & 24$\times$300s = 2h00m \\ \hline
CD-35 2722 & \begin{tabular}{@{}l}CD-35 2722A, \\CD-35 2722B\end{tabular} & 22.36$\pm$0.01$^{(2)}$ & ABDMG (99.9\%) & 149$^{+51}_{-19}$$^{(9)}$ & \begin{tabular}{@{}l}M1$^{(30)}$, \\L4$^{(31)}$\end{tabular} & \begin{tabular}{@{}l}7.86$\pm$0.03, \\13.63$\pm$0.11$^{(31)}$\end{tabular} & \begin{tabular}{@{}l}7.03$\pm$0.05, \\12.01$\pm$0.07$^{(31)}$\end{tabular} & \begin{tabular}{@{}l}419$\pm$52, \\31$\pm$8$^{(31)}$\end{tabular} & 3.14/67$\pm$4$^{(31)}$ & 2022 12 31 & K2166 & No & 0.4 & 8$\times$300s = 0h40m \\
 &  &  &  &  &  &  &  &  &  & 2023 01 03 & K2166 & No & 0.4 & 24$\times$300s = 2h00m \\
 &  &  &  &  &  &  &  &  &  & 2023 01 31 & J1226 & No & 0.4 & 7$\times$300s = 0h35m \\ \cline{15-15} 
 &  &  &  &  &  &  &  &  &  &  &  &  &  & 3h15m \\ \hline
AB Pic & AB Pic b & 50.14$\pm$0.03$^{(2)}$ & CAR (99.7\%) & 13.3$^{+1.1}_{-0.6}$$^{(32)}$ & L0-1$^{(33)}$ & 16.18$\pm$0.10$^{(34)}$ & 14.14$\pm$0.08$^{(34)}$ & 10$\pm$1$^{(35)}$ & 5.4/273$\pm$2$^{(34)}$ & 2022 11 01 & K2166 & Yes & 0.4 & 22$\times$300s = 1h50m \\
 &  &  &  &  &  &  &  &  &  & 2022 11 02 & K2166 & Yes & 0.4 & 12$\times$300s = 1h00m \\
 &  &  &  &  &  &  &  &  &  & 2022 11 03 & K2166 & Yes & 0.4 & 16$\times$300s = 1h20m \\
 &  &  &  &  &  &  &  &  &  & 2022 11 04 & K2166 & Yes & 0.4 & 14$\times$300s = 1h10m \\ \cline{15-15} 
 &  &  &  &  &  &  &  &  &  &  &  &  &  & 5h20m \\ \hline
2MASS J06244595-4521548 & 2M 0624 & 12.19$\pm$0.05$^{(2)}$ & ARG (93.5\%) & 40-50$^{(36)}$ & L6.5$^{(15)}$ & 14.48$\pm$0.03$^{(7)}$ & 12.60$\pm$0.03$^{(7)}$ & - & - & 2023 02 27 & K2166 & No & 0.4 & 15$\times$300 = 1h15m \\ \hline
SSSPM J0829-1309 & 2M 0829 & 11.68$\pm$0.02$^{(2)}$ & FIELD (99.9\%) & - & L2$^{(37)}$ & 12.80$\pm$0.03$^{(7)}$ & 11.30$\pm$0.02$^{(7)}$ & - & - & 2023 01 03 & K2166 & No & 0.4 & 12$\times$300s = 1h00m \\ \hline
2MASS J08354256-0819237 & 2M 0835 & 7.23$\pm$0.01$^{(2)}$ & FIELD (95.1\%) & - & L6.5$^{(15)}$ & 13.17$\pm$0.02$^{(7)}$ & 11.136$\pm$0.02$^{(7)}$ & - & - & 2023 01 01 & K2166 & No & 0.4 & 4$\times$300s = 0h20m \\ \hline
HR 3549 & HR 3549B & 94.8$\pm$0.3$^{(2)}$ & FIELD (92.1\%) & 100-150$^{(38)}$ & L0$^{(38)}$ & 15.94$\pm$0.06$^{(38)}$ & - & 40-50$^{(38)}$ & 0.9/80$^{(39)}$ & 2023 01 01 & K2166 & Yes & 0.2 & 26$\times$300s = 2h10m \\ \hline
2MASS J08561384-1342242 & 2M 0856 & 53.8$\pm$0.4$^{(2)}$ & FIELD (96.2\%) & - & M8$^{(23)}$ & 13.60$\pm$0.03$^{(7)}$ & 12.49$\pm$0.02$^{(7)}$ & - & - & 2023 03 04 & K2166 & No & 0.4 & 18$\times$300s = 1h30m \\ \hline
2MASSI J0953212-101420 & 2M 0953 & 35.7$\pm$0.3$^{(2)}$ & FIELD (63.5\%) & - & L0.0$^{(29)}$ & 13.47$\pm$0.03$^{(7)}$ & 12.14$\pm$0.02$^{(7)}$ & - & - & 2023 03 04 & K2166 & No & 0.4 & 18$\times$300s = 1h30m \\ \hline
Luhman 16 & \begin{tabular}{@{}l}Luhman 16A, \\Luhman 16B\end{tabular} & 2.02$\pm$0.15$^{(40)}$ & ARG (94.1\%) & 40-50$^{(36)}$ & \begin{tabular}{@{}l}L7.5, \\T0.5$^{(41)}$\end{tabular} & \begin{tabular}{@{}l}11.53$\pm$0.04, \\11.22$\pm$0.04$^{(41)}$\end{tabular} & \begin{tabular}{@{}l}9.44$\pm$0.07, \\9.73$\pm$0.09$^{(41)}$\end{tabular} & \begin{tabular}{@{}l}33.5$\pm$0.3, \\28.6$\pm$0.3$^{(42)}$\end{tabular} & 1.5/3.12$\pm$0.25$^{(40)}$ & 2022 12 31 & K2166 & No & 0.2 & 4$\times$300s = 0h20m \\
 &  &  &  &  &  &  &  &  &  & 2023 01 01 & K2166 & No & 0.4 & 4$\times$300s = 0h20m \\
 &  &  &  &  &  &  &  &  &  & 2023 01 01 & J1226 & No & 0.4 & 4$\times$300s = 0h20m \\ \cline{15-15} 
 &  &  &  &  &  &  &  &  &  &  &  &  &  & 1h00m \\ \hline
TWA 28 & TWA 28 & 59.2$\pm$0.4$^{(2)}$ & TWA (99.6\%) & 10$\pm$3$^{(9)}$ & M8.3$^{(43)}$ & 13.03$\pm$0.02$^{(7)}$ & 11.89$\pm$0.02$^{(7)}$ & - & - & 2023 03 03 & K2166 & No & 0.4 & 14$\times$300s = 1h10m \\ \hline
CD-33 7795 & \begin{tabular}{@{}l}TWA 5A, \\TWA 5B\end{tabular} & 49.6$\pm$0.1$^{(2)}$ & TWA (99.9\%) & 10$\pm$3$^{(9)}$ & \begin{tabular}{@{}l}M1.5, \\M8.5$^{(44)}$\end{tabular} & \begin{tabular}{@{}l}8.40$\pm$0.07, \\12.6$\pm$0.2$^{(18)}$\end{tabular} & \begin{tabular}{@{}l}7.39$\pm$0.04, \\11.4$\pm$0.2$^{(18)}$\end{tabular} & \begin{tabular}{@{}l}$\sim$419$^{(45)}$, \\25$^{+120}_{-20}$$^{(46)}$\end{tabular} & 1.9/86$\pm$4$^{(46)}$ & 2023 02 26 & K2166 & Yes & 0.2 & 16$\times$300s = 1h20m \\ \hline
2MASS J11553952-3727350 & 2M 1155 & 11.80$\pm$0.02$^{(2)}$ & FIELD (99.9\%) & - & L2$^{(47)}$ & 12.81$\pm$0.02$^{(7)}$ & 11.46$\pm$0.02$^{(7)}$ & - & - & 2023 02 01 & K2166 & No & 0.4 & 4$\times$300s = 0h20m \\
 &  &  &  &  &  &  &  &  &  & 2023 02 01 & J1226 & No & 0.4 & 2$\times$300s = 0h10m \\ \cline{15-15} 
 &  &  &  &  &  &  &  &  &  &  &  &  &  & 0h30m \\ \hline
2MASS J12003792-7845082 & 2M 1200 & 101.6$\pm$0.7$^{(2)}$ & EPSC (99.8\%) & 3.7$^{+4.6}_{-1.4}$$^{(48)}$ & M6$^{(49)}$ & 12.52$\pm$0.02$^{(7)}$ & 11.60$\pm$0.02$^{(7)}$ & - & - & 2023 03 04 & K2166 & No & 0.4 & 8$\times$300s = 0h40m \\ \hline
HIP 64892 & HIP 64892B & 119.6$\pm$0.7$^{(2)}$ & LCC (74.0\%) & 15$\pm$3$^{(50)}$ & M9$^{(51)}$ & 14.85$\pm$0.15$^{(51)}$ & 13.50$\pm$0.17$^{(51)}$ & 33$\pm$4$^{(51)}$ & 1.27/159$\pm$12$^{(51)}$ & 2023 02 26 & K2166 & Yes & 0.2 & 12$\times$300s = 1h00m \\
 &  &  &  &  &  &  &  &  &  & 2023 03 03 & K2166 & Yes & 0.2 & 5$\times$300s = 0h25m \\
 &  &  &  &  &  &  &  &  &  & 2023 03 04 & K2166 & Yes & 0.2 & 18$\times$300s = 1h30m \\ \cline{15-15} 
 &  &  &  &  &  &  &  &  &  &  &  &  &  & 2h55m \\ \hline
YSES 1 & \begin{tabular}{@{}l}YSES 1b, \\YSES 1c\end{tabular} & 94.2$\pm$0.1$^{(2)}$ & LCC (99.1\%) & 15$\pm$3$^{(50)}$ & \begin{tabular}{@{}l}L0$^{(52)}$, \\L7.5$^{(53)}$\end{tabular} & \begin{tabular}{@{}l}15.73$\pm$0.38$^{(52)}$, \\21.50$\pm$0.13$^{(53)}$\end{tabular} & \begin{tabular}{@{}l}14.70$\pm$0.14$^{(52)}$, \\18.13$\pm$0.13$^{(53)}$\end{tabular} & \begin{tabular}{@{}l}14$\pm$3$^{(52)}$, \\6$\pm$1$^{(53)}$\end{tabular} & \begin{tabular}{@{}l}1.71/162$^{(52)}$, \\3.37/320$^{(53)}$\end{tabular} & 2023 02 26 & K2166 & Yes & 0.2 & 22$\times$600s = 3h40m \\
 &  &  &  &  &  &  &  &  &  & 2023 02 27 & K2166 & Yes & 0.2 & 10$\times$600s = 1h40m \\ \cline{15-15} 
 &  &  &  &  &  &  &  &  &  &  &  &  &  & 5h20m \\ \hline
2MASS J14252798-3650229 & 2M 1425 & 11.85$\pm$0.04$^{(2)}$ & ABDMG (99.9\%) & 149$^{+51}_{-19}$$^{(9)}$ & L4$^{(23)}$ & 13.75$\pm$0.03$^{(7)}$ & 11.81$\pm$0.03$^{(7)}$ & - & - & 2023 02 01 & K2166 & No & 0.4 & 12$\times$300s = 1h00m \\
 &  &  &  &  &  &  &  &  &  & 2023 02 01 & J1226 & No & 0.4 & 4$\times$300s = 0h20m \\ \cline{15-15} 
 &  &  &  &  &  &  &  &  &  &  &  &  &  & 1h20m \\ \hline
GQ Lup & \begin{tabular}{@{}l}GQ Lup A, \\GQ Lup b\end{tabular} & 154.1$\pm$0.7$^{(2)}$ & UCL (99.4\%) & 16$\pm$2$^{(50)}$ & \begin{tabular}{@{}l}K7$^{(54)}$, \\L1$^{(55)}$\end{tabular} & \begin{tabular}{@{}l}8.69$\pm$0.04, \\14.90$\pm$0.11$^{(18)}$\end{tabular} & \begin{tabular}{@{}l}7.10$\pm$0.02, \\13.34$\pm$0.13$^{(18)}$\end{tabular} & \begin{tabular}{@{}l}1079$\pm$52$^{(56)}$, \\$\sim$30$^{(57)}$\end{tabular} & 0.7/105$^{(18)}$ & 2023 02 27 & K2166 & Yes & 0.2 & 61$\times$180s = 3h03m \\ \hline
ROXs 12 & \begin{tabular}{@{}l}ROXs 12A, \\ROXs 12b\end{tabular} & 138.6$\pm$0.3$^{(2)}$ & USCO (97.6\%) & 10$\pm$3$^{(50)}$ & \begin{tabular}{@{}l}M0, \\L0$^{(58)}$\end{tabular} & \begin{tabular}{@{}l}10.28$\pm$0.02, \\15.82$\pm$0.03$^{(58)}$\end{tabular} & \begin{tabular}{@{}l}9.10$\pm$0.03, \\14.14$\pm$0.03$^{(58)}$\end{tabular} & \begin{tabular}{@{}l}911$\pm$84, \\16$\pm$4$^{(59)}$\end{tabular} & 1.75/210$\pm$20$^{(59)}$ & 2023 03 03 & K2166 & No & 0.4 & 2$\times$450s = 0h15m \\
 &  &  &  &  &  &  &  &  &  & 2023 03 04 & K2166 & No & 0.4 & 22$\times$600s = 3h40m \\ \cline{15-15} 
 &  &  &  &  &  &  &  &  &  &  &  &  &  & 3h55m \\ \hline
PZ Tel & PZ Tel B & 47.25$\pm$0.05$^{(2)}$ & $\beta$PMG (97.2\%) & 24$\pm$3$^{(9)}$ & M7$^{(60)}$ & 12.26$\pm$0.14$^{(61)}$ & 11.42$\pm$0.15$^{(61)}$ & 27$^{+25}_{-9}$$^{(62)}$ & 0.48/$\sim$25$^{(60)}$ & 2022 11 02 & K2166 & Yes & 0.2 & 8$\times$300s = 0h40m \\
 & \begin{tabular}{@{}l}PZ Tel A, \\PZ Tel B\end{tabular} &  &  &  & \begin{tabular}{@{}l}G9$^{(63)}$, \\M7$^{(60)}$\end{tabular} & \begin{tabular}{@{}l}6.86$\pm$0.02$^{(7)}$, \\12.26$\pm$0.14$^{(61)}$\end{tabular} & \begin{tabular}{@{}l}6.38$\pm$0.02$^{(7)}$, \\11.42$\pm$0.15$^{(61)}$\end{tabular} & \begin{tabular}{@{}l}1184$\pm$31$^{(64)}$, \\27$^{+25}_{-9}$$^{(62)}$\end{tabular} & 0.48/$\sim$25$^{(60)}$ & 2022 11 03 & K2166 & Yes & 0.2 & 22$\times$150s = 0h55m \\ \cline{15-15} 
 &  &  &  &  &  &  &  &  &  &  &  &  &  & 1h35m \\ \hline
36 systems & \begin{tabular}{@{}l}49 on-slit, \\19 companions\end{tabular} &  &  &  &  &  &  &  &  &  &  &  &  & 

\end{longtable}

\tablefoot{As a consequence of the degeneracy between the ages and masses of isolated, field sub-stellar objects, these parameters are not listed. The BANYAN $\Sigma$ association \citep{Gagne_ea_2018} with the highest membership probability (in parentheses) is shown. The full names of these identifiers are field (FIELD), Tucana-Horologium association (THA), AB Doradus (ABDMG), Taurus (TAU), $\beta$ Pictoris ($\beta$PMG), Carina (CAR), Argus (ARG), TW Hya (TWA), $\epsilon$ Chamaeleontis (EPSC), Lower Centaurus Crux (LCC), Upper Centaurus Lupus (UCL), and Upper Scorpius (USCO).}

\tablebib{
    (1)~\citet{Gagne_ea_2018};
    (2) \citet{Gaia_2020};
    (3) \citet{Mesa_ea_2020};
    (4) \citet{Garcia_ea_2017};
    (5) \citet{Nielsen_ea_2012};
    (6) \citet{Reid_ea_2000};
    (7) \citet{2MASS_2003};
    (8) \citet{Riedel_ea_2014};
    (9) \citet{Bell_ea_2015};
    (10) \citet{Delorme_ea_2013};
    (11) \citet{Riaz_ea_2006};
    (12) \citet{Bowler_ea_2013};
    (13) \citet{Cruz_ea_2009};
    (14) \citet{Artigau_ea_2015};
    (15) \citet{Burgasser_ea_2006};
    (16) \citet{Kenyon_ea_1995};
    (17) \citet{Luhman_ea_2009};
    (18) \citet{Patience_ea_2012};
    (19) \citet{Xuan_ea_2020};
    (20) \citet{Luhman_ea_2006};
    (21) \citet{Itoh_ea_2005};
    (22) \citet{Monin_ea_2010};
    (23) \citet{Gagne_ea_2015};
    (24) \citet{Cruz_ea_2003};
    (25) \citet{Chilcote_ea_2017};
    (26) \citet{Lacour_ea_2021};
    (27) \citet{GRAVITY_2020};
    (28) \citet{Geballe_ea_2002};
    (29) \citet{Bardalez_Gagliuffi_ea_2014};
    (30) \citet{Torres_ea_2006};
    (31) \citet{Wahhaj_ea_2011};
    (32) \citet{Booth_ea_2021};
    (33) \citet{Bonnefoy_ea_2010};
    (34) \citet{Chauvin_ea_2005};
    (35) \citet{Palma_Bifani_ea_2023};
    (36) \citet{Zuckerman_2018};
    (37) \citet{Scholz_ea_2002};
    (38) \citet{Mesa_ea_2016};
    (39) \citet{Mawet_ea_2015};
    (40) \citet{Luhman_ea_2013};
    (41) \citet{Burgasser_ea_2013};
    (42) \citet{Lazorenko_ea_2018};
    (43) \citet{Faherty_ea_2016};
    (44) \citet{Lowrance_ea_1999};
    (45) \citet{Faherty_ea_2010};
    (46) \citet{Neuhauser_ea_2010};
    (47) \citet{Gizis_2002};
    (48) \citet{Murphy_ea_2013};
    (49) \citet{Schutte_ea_2020};
    (50) \citet{Pecaut_ea_2016};
    (51) \citet{Cheetham_ea_2018};
    (52) \citet{Bohn_ea_2020a};
    (53) \citet{Bohn_ea_2020b};
    (54) \citet{Herbig_ea_1977};
    (55) \citet{Lavigne_ea_2009};
    (56) \citet{MacGregor_ea_2017};
    (57) \citet{Stolker_ea_2021};
    (58) \citet{Bowler_ea_2017};
    (59) \citet{Kraus_ea_2014};
    (60) \citet{Maire_ea_2016};
    (61) \citet{Biller_ea_2010};
    (62) \citet{Franson_ea_2023};
    (63) \citet{Stolker_ea_2020};
    (64) \citet{Jenkins_ea_2012}
}

}
}
\end{landscape}

\section{PT profile parameterisation} \label{app:PT}
\subsection{Definition} \label{app:PT_def}
Here, we describe the method used to compute the general difference matrix $\vec{D}_3$, following the general, non-uniform P-splines formalism of \citet{Li_ea_2022}. As described in Sect. \ref{sect:PT}, the spline coefficient vector $\vec{C}$ is obtained with \texttt{scipy} in $\log(T)$-$\log(P)$ space. We employ \texttt{scipy}'s "not-a-knot" method where the third derivative remains continuous over the first and final two segments. Effectively, this removes the second and second-to-last knots from the sequence. However, these knots still influence the spline coefficients in $\vec{C}$ and their oscillations are still penalised as a result. With a clamped boundary and defining $N$ points in log pressure space, the knots $t_i$ are assigned as:
\begin{align}
\begin{tabular}{|c|c|c|c|c|c|}
$t_1$, $t_2$, $t_3$, $t_4$ & $t_5$       & $t_6$       & ... & $t_{K-4}$       & $t_{K-3}$, $t_{K-2}$, $t_{K-1}$, $t_{K}$ \\ \hline
$\log(P_1)$                & $\log(P_3)$ & $\log(P_4)$ & ... & $\log(P_{N-2})$ & $\log(P_N)$
\end{tabular}.
\end{align}
As described by \citet{Li_ea_2022}, the lag-3, lag-2, and lag-1 differences give the diagonal weight matrices $\vec{W}_1$, $\vec{W}_2$, $\vec{W}_3$, respectively. The boundary knots are successively turned "inactive", so that the weight matrices become
\begin{align}
    \vec{W}_1 &= \frac{1}{3}\begin{bmatrix}
        t_5 - t_2 & & & & & \\
        & t_6 - t_3 & & & & \\
        & & t_7 - t_4 & & & \\
        & & & \ddots & & \\
        & & & & t_{K-2}-t_{K-5} & \\
        & & & & & t_{K-1}-t_{K-4} \\
    \end{bmatrix},  \\
    \vec{W}_2 &= \frac{1}{2}\begin{bmatrix}
        t_5 - t_3 & & & & & \\
        & t_6 - t_4 & & & & \\
        & & t_7 - t_5 & & & \\
        & & & \ddots & & \\
        & & & & t_{K-3}-t_{K-5} & \\
        & & & & & t_{K-2}-t_{K-4} \\
    \end{bmatrix},  \\
    \vec{W}_3 &= \frac{1}{1}\begin{bmatrix}
        t_5 - t_4 & & & & & \\
        & t_6 - t_5 & & & & \\
        & & t_7 - t_6 & & & \\
        & & & \ddots & & \\
        & & & & t_{K-3}-t_{K-5} & \\
        & & & & & t_{K-3}-t_{K-4} \\
    \end{bmatrix}.
\end{align}
The fundamental difference matrix is defined as
\begin{align}
    \vec{\Delta} &= \begin{bmatrix}
        -1 & 1 & & & \\
        & -1 & 1 & & \\
        & & \ddots & \ddots & \\
        & & & -1 & 1 \\
    \end{bmatrix}, 
\end{align}
where its exact dimensions alter, depending on the calculation performed. These matrices are used to compute the variation between coefficients in the vector $\vec{C}$ as $\vec{\Delta}\vec{C}$. Consecutively performing this difference computation and weighting by the inverse of $\vec{W}_i$ results in the first-, second- and third-order general differences:
\begin{align}
    \vec{C}_1 &= \vec{W}_1^{-1}\vec{\Delta} \vec{C}, \\
    \vec{C}_2 &= \vec{W}_2^{-1}\vec{\Delta} \vec{W}_1^{-1}\vec{\Delta} \vec{C}, \\
    \vec{C}_3 &= \underbrace{\vec{W}_3^{-1}\vec{\Delta} \vec{W}_2^{-1}\vec{\Delta} \vec{W}_1^{-1}\vec{\Delta}}_{\vec{D}_3} \vec{C}, 
\end{align}
where $\vec{D}_3$ is the third-order general difference matrix. The third-order general difference penalty is then calculated as the squared norm of the general difference:
\begin{align}
    \mathrm{PEN}^{(3)}_\mathrm{gps}=\|\vec{D}_3\vec{C}\|^2.
\end{align}

\subsection{Retrieval of synthetic spectrum} \label{app:PT_synth}
To verify the applicability of the updated PT parameterisation, we use a synthetic spectrum to evaluate the retrieval of its input parameters. As the input PT profile, we employ a Sonora Bobcat profile \citep{Marley_ea_2021} with an effective temperature of $T_\mathrm{eff}=1400\ \mathrm{K}$, surface gravity $\log\ \textit{g}=5.25$, solar metallicity and solar C/O-ratio \citep{Cushing_ea_2008,Tremblin_ea_2016,Charnay_ea_2018,Lueber_ea_2022}. Furthermore, we adopt $R=0.8\ R_\mathrm{Jup}$, $\log\ \mathrm{^{12}CO}=-3.3$, $\log\ \mathrm{^{13}CO}=-5.5$, $\log\ \mathrm{H_2O}=-3.6$, $\log\ \mathrm{CH_4}=-4.9$, $\log\ \mathrm{NH_3}=-6.0$, $\varepsilon_\mathrm{limb}=0.65$, $\textit{v}\sin i=41.0\ \mathrm{km\ s^{-1}}$, and $\textit{v}_\mathrm{rad}=22.5\ \mathrm{km\ s^{-1}}$ \citep{Basri_ea_2000,Mohanty_ea_2003,Zapatero_Osorio_ea_2006}. After generating a \texttt{pRT} spectrum with these parameters, the spectrum is convolved down to the CRIRES$^+$ spectral resolving power ($R\sim$\,$100\,000$). Similar to the observed spectrum, pixels are masked where deep tellurics exist. Gaussian noise is added to the synthetic spectrum by using the flux uncertainties of the observed spectrum of DENIS J0255. For the retrieval on this synthetic spectrum, we use the same priors as the fiducial model (see Table \ref{tab:params}) except for the cloud and GP parameters, which are not retrieved. 

\begin{figure*}[h!]
    \centering
    \includegraphics[width=17cm]{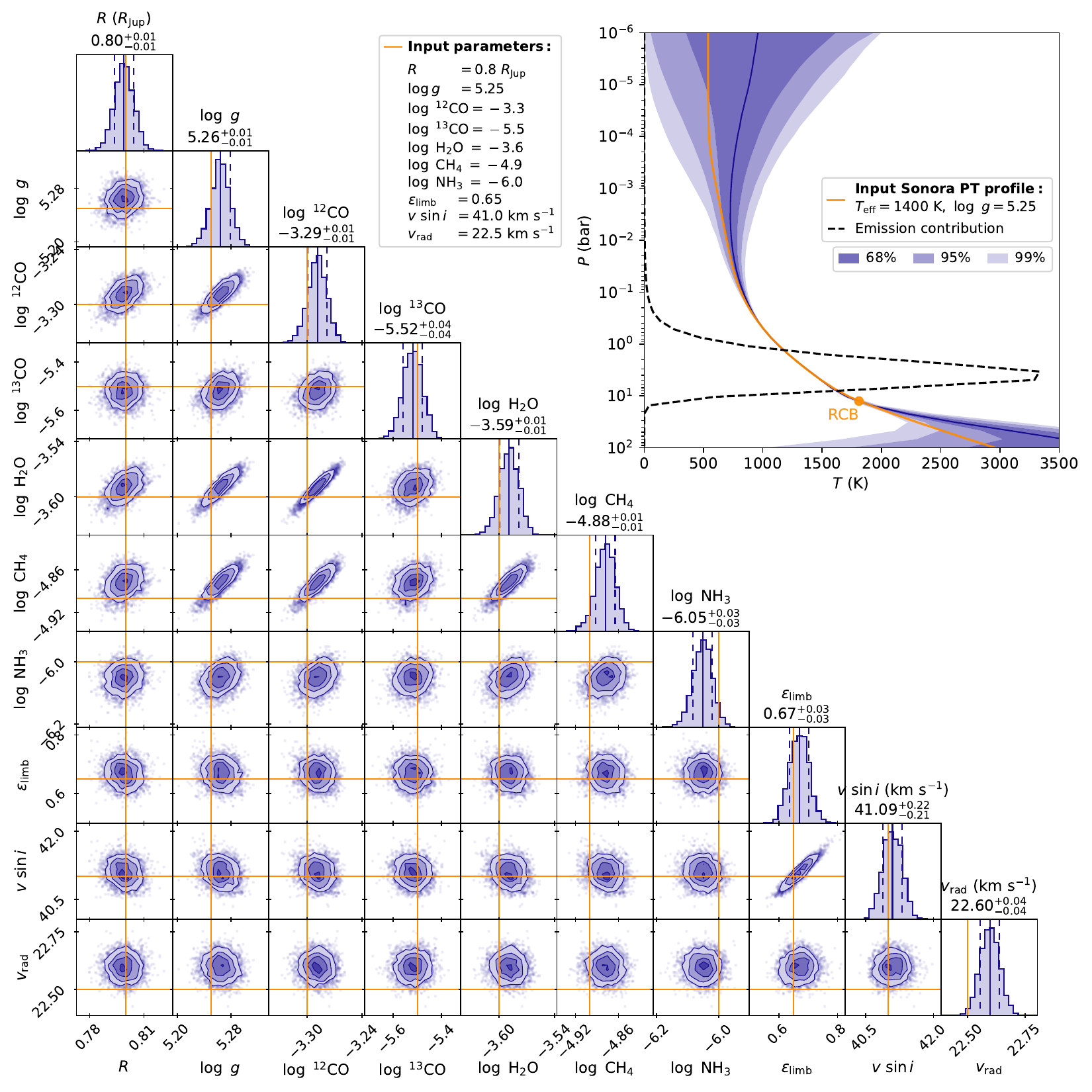}
    \caption{Posterior distributions retrieved from the synthetic spectrum. The orange lines denote the input parameters and PT profile, which are compatible with the retrieval within the $95\%$ credible region. Following the Schwarzschild criterion, the radiative-convective boundary (RCB) is indicated with a point near $10\ \mathrm{bar}$ and $1800\ \mathrm{K}$.}
    \label{fig:synthetic_corner}
\end{figure*}

The results of the retrieval are presented in Fig. \ref{fig:synthetic_corner}. The retrieved posterior distributions are compatible with the input parameters, indicated by orange lines, within the $95\%$ confidence interval. As a known deficiency of the \texttt{MultiNest} algorithm, the sampled posteriors are likely under-dispersed (e.g. \citealt{Buchner_2016, Ardevol_Martinez_ea_2022, Chubb_ea_2022, Vasist_ea_2023, Latouf_ea_2023}), which might explain the offset of the retrieved radial velocity. Below the photosphere, we observe that the retrieved PT envelope is somewhat steeper than the input Sonora Bobcat profile. The synthetic spectrum provides negligible information about this region of the atmosphere, thus resulting in a PT envelope that covers the given prior, albeit smoothened by the third-order general difference penalty. Inside the photosphere, the Sonora Bobcat PT profile is almost perfectly recovered. Therefore, we expect that the updated PT profile parameterisation is well applicable to the observed CRIRES$^+$ spectrum and should be capable of constraining the atmospheric properties of DENIS J0255. 

\section{Gaussian Processes}\label{app:GPs}
\begin{figure*}[h!]
    \centering
    \includegraphics[width=17cm]{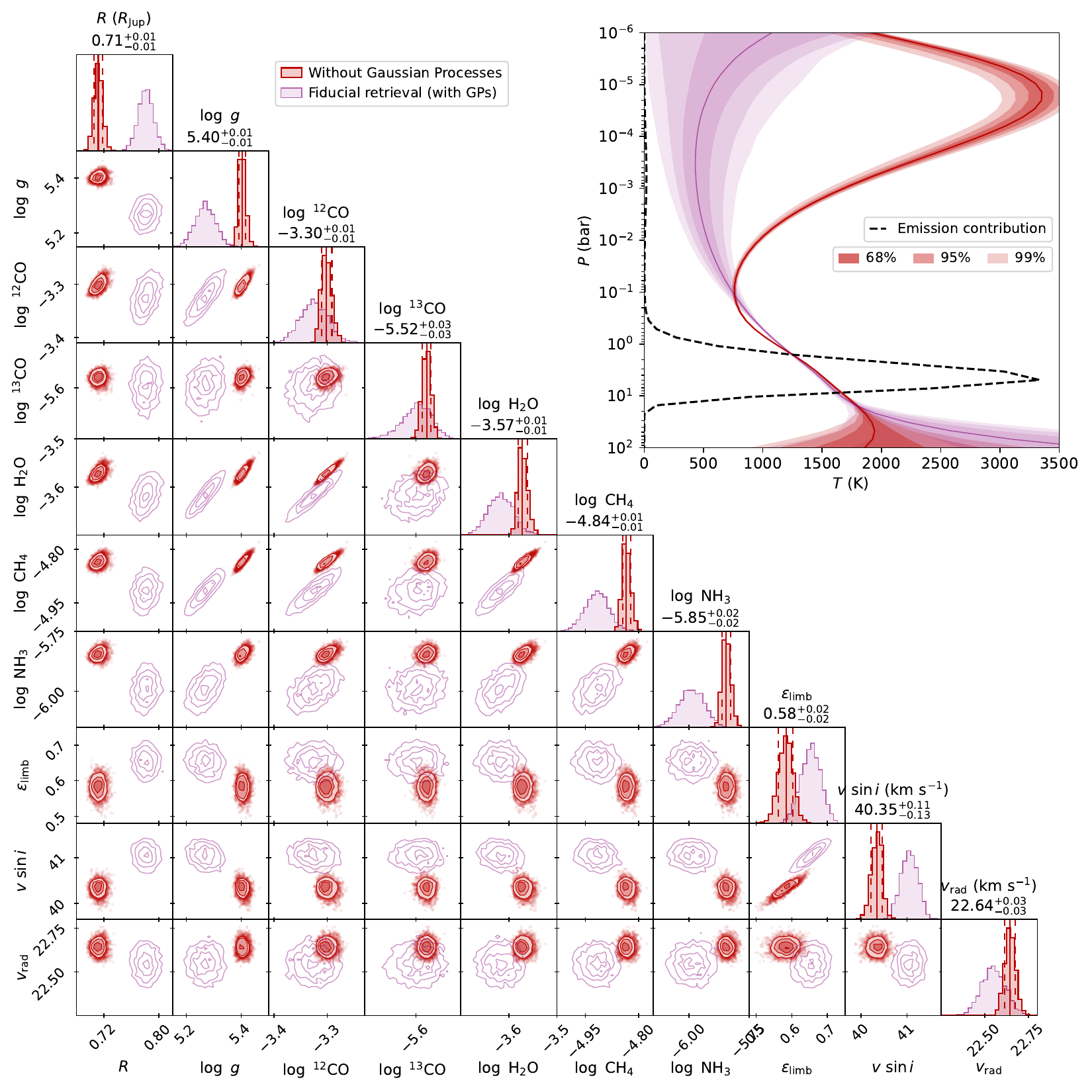}
    \caption{Comparison between the retrieval results of a model without Gaussian Processes (red) and the fiducial model (pink). The values shown above each 1D histogram provide the median and $68\%$ confidence interval of the retrieval without Gaussian Processes. The thermal inversion in the \textit{upper right panel} and the narrow and biased posteriors in the \textit{lower left panels} are the result of overfitting when covariance is not modelled. }
    \label{fig:wo_GPs_corner}
\end{figure*}

To demonstrate the importance of modelling the off-diagonal covariance elements, we carry out a retrieval without Gaussian Processes. Apart from the GP parameters, we utilise the same priors as the fiducial model (see Table \ref{tab:params}). We note that the optimal covariance scaling parameter $\tilde{s}^2$ is still calculated, but is different from the fiducial model as it only scales the flux-uncertainties. The posterior distributions of the retrieval without GPs is shown in red in Fig. \ref{fig:wo_GPs_corner} and compared to the fiducial retrieval results in pink. Since introducing GPs effectively smoothens out the likelihood landscape, the results without GPs show more narrow posteriors and smaller uncertainties. Moreover, the retrieved values are substantially biased if covariance is not accounted for. For example, a large temperature inversion is constrained at $\sim$\,$10^{-4}\ \mathrm{bar}$ in order to make the deepest H$_2$O lines more shallow at the K-band edges. This inversion is the result of overfitting as these lines are also well-reproduced by the fiducial model (see Fig. \ref{fig:H2O_line_list}). Similarly, we find a radius and surface gravity discrepant from the results obtained in the fiducial retrieval. In spite of some minor offsets in the abundances, the corresponding carbon-to-oxygen ratio $\mathrm{C/O}=0.671^{+0.001}_{-0.001}$, metallicity $[\mathrm{C/H}]=0.05^{+0.01}_{-0.01}$, and carbon isotope ratio $\mathrm{^{12}C/^{13}C}=164^{+13}_{-12}$ are still compatible with those found with the fiducial model. The discrepancies between retrievals with and without covariance modelling are expected to be smaller for spectra that experience less rotational broadening than DENIS J0255 ($\textit{v}\sin i\sim$\,$41\ \mathrm{km\ s^{-1}}$), considering the reduced correlation between adjacent pixels.

\end{appendix}

\end{document}